\documentclass[twocolumn,superscriptaddress,floatfix,preprintnumbers,prd,nofootinbib]{revtex4-2}
\usepackage[colorlinks=true,breaklinks=true]{hyperref}
\usepackage{multirow}
\usepackage[utf8]{inputenc}
\hypersetup{urlcolor=Blue,
	    citecolor=Blue,
	    linkcolor=Blue}
\usepackage{url}
\usepackage{mathtools}
\usepackage{amssymb}
\usepackage[usenames,dvipsnames]{xcolor}

\usepackage{orcidlink}
\usepackage{enumitem}
\usepackage{slashed}
\definecolor{darkblue}{RGB}{0,0,180}

\newcommand{\bP}{{\bf P}}

\newcommand{\ba}{{\bf a}}
\newcommand{\bb}{{\bf b}}

\newcommand{\bn}{{\bf n}}
\newcommand{\bk}{{\bf k}}
\newcommand{\bq}{{\bf q}}

\newcommand{\bl}{{\bf l}}

\newcommand{\bp}{{\bf p}}

\newcommand{\RNS}{R_{\rm NS}}
\newcommand{\Rprog}{R_{\rm prog}}
\newcommand{\edit}[1]{{\color{darkblue}#1}}

\newcommand{\exclude}[1]{{}}
\long\def\exclude#1{}

\long\def\exclude#1{}

\bibpunct{[}{]}{,}{n}{}{,}

\begin{document}

\title{Supernova production of axion-like particles coupling to electrons, reloaded}

\author{Damiano F.\ G.\ Fiorillo \orcidlink{0000-0003-4927-9850}}
\affiliation{Deutsches Elektronen-Synchrotron DESY,
Platanenallee 6, 15738 Zeuthen, Germany}

\author{Tetyana Pitik \orcidlink{0000-0002-9109-2451}} 
\affiliation{Department of Physics, University of California, Berkeley, Berkeley, CA 94720, USA}
\affiliation{Institute for Gravitation and the Cosmos, The Pennsylvania State University, University Park PA 16802, USA}

\author{Edoardo Vitagliano \orcidlink{0000-0001-7847-1281}} 
\affiliation{Dipartimento di Fisica e Astronomia, Universit\`a degli Studi di Padova, Via Marzolo 8, 35131 Padova, Italy}
\affiliation{Istituto Nazionale di Fisica Nucleare (INFN), Sezione di Padova, Via Marzolo 8, 35131 Padova, Italy}

\date{March 26, 2025, \edit{Post-publication updates in blue, \today}}

\begin{abstract}
    We revisit the production of axion-like particles (ALPs) coupled to electrons at tree-level in a relativistic plasma. We explicitly demonstrate the equivalence between pseudoscalar and derivative couplings, incorporate previously neglected processes for the first time---namely, semi-Compton production ($\gamma e^-\rightarrow a e^-$) and pair annihilation ($e^+e^-\rightarrow a\gamma$)---and derive analytical expressions for the bremsstrahlung ($e^- N\to e^- N a$) production rate, enabling a more computationally efficient evaluation of the ALP flux. Additionally, we assess uncertainties in the production rate arising from electron thermal mass corrections, electron-electron Coulomb interactions, and the Landau-Pomeranchuk-Migdal effect. The ALP emissivity is made available in a public repository as a function of the ALP mass, the temperature, and the electron chemical potential of the plasma. Finally, we examine the impact of ALP production and subsequent decays on
    astrophysical observables, deriving the leading bounds on ALPs coupling to electrons.
    At small couplings, the dominant constraints come from the previously neglected decay $a\to e^+ e^-\gamma$, except for a region of fireball formation where SN~1987A X-ray observations offer the best probe. At large couplings, bounds are dominated by the energy deposition argument, with a recently developed new prescription for the trapping regime.
\end{abstract}

\maketitle

\tableofcontents

\section{Introduction}
Astrophysical transients provide optimal targets for discovering novel feebly interacting particles with masses in the MeV-GeV range. The extreme temperatures and densities reached in the neutron star (NS) remnants of core-collapse supernovae (SNe)---for a sample of the vast literature, see e.g.~\cite{Oberauer:1993yr,Davidson:2000hf,Giannotti:2010ty,Kazanas:2014mca,Chang:2016ntp,Jaeckel:2017tud,Chang:2018rso,Brdar:2020quo,Croon:2020lrf,Camalich:2020wac,Lucente:2020whw,Caputo:2021rux,Caputo:2022mah,Caputo:2022rca,Ferreira:2022xlw,Hoof:2022xbe,Lella:2022uwi,Fiorillo:2022cdq,Akita:2022etk,Caputo:2022rca,Diamond:2023scc,Lella:2023bfb,Fiorillo:2023cas, Fiorillo:2023ytr,Carenza:2023old,Akita:2023iwq,Lella:2024dmx,Fiorillo:2024upk,Telalovic:2024cot,Benabou:2024jlj,Alda:2024cxn}---as well as binary mergers~\cite{Diamond:2021ekg,Fiorillo:2022piv,Diamond:2023cto,Dev:2023hax} and other rare astrophysical events such as hypernovae~\cite{Caputo:2021kcv}, allow many extensions of the Standard Model (SM) of particle physics to be tested via multimessenger observations.

To assess the impact of new particles on astrophysical observables, simplified models are often employed. For example, axion-like particles (hereafter ALPs) are pseudoscalars that generalize the QCD axion and can couple to SM particles in different ways (see e.g.~\cite{DiLuzio:2020wdo} for a review). In this work, we focus on their coupling to electrons, described by the Lagrangian:
 \begin{equation}
    \mathcal{L}_{\rm a}=\mathcal{L}_{\rm SM}+\frac{1}{2}\partial_{\mu} a \partial^{\mu} a-\frac{1}{2}m_a^2 a^2 +g_a a\bar{e}\gamma_5 e.
\end{equation}
 If $m_a\gtrsim 1\,\rm MeV$, these ALPs cannot be produced in the relatively cold cores of red-giant branch stars and white dwarfs,
 which otherwise provide the strongest constraints on $g_a$~\cite{Raffelt:1985nj,Corsico:2012ki,Corsico:2012sh,MillerBertolami:2014rka,Battich:2016htm} (see also Ref.~\cite{Bottaro:2023gep} for the scalar case). Likewise, cosmological bounds are competitive only if the reheating temperature is large enough to generate a large ALP population in the early universe~\cite{Depta:2020zbh,Langhoff:2022bij}. 
 Core-collapse SNe thus remain a natural probe for ALPs in the mass range $1\,{\rm MeV}\lesssim m_a\lesssim 1\,\rm GeV$, making a precise determination of ALP emission via their coupling to electrons at SN-relevant temperatures and densities crucial. This serves as the key ingredient for deriving constraints, similar to those previously obtained for ALP-photon couplings, based on the cooling and X-ray/gamma-ray observations of SN~1987A, the diffuse gamma-ray flux, and the energetics of low-energy SNe~\cite{Lucente:2020whw,Calore:2020tjw, Caputo:2021rux,Caputo:2022mah,Hoof:2022xbe,
 Diamond:2023scc}. Additionally, ALPs coupling to electrons can contribute to a positron excess, leading to constraints similar to those applied in the context of dark photons~\cite{Kazanas:2014mca,DeRocco:2019njg,Calore:2021lih}. In recent years, several efforts have been made to compute the ALP production spectrum~\cite{Carenza:2021pcm,Ferreira:2022xlw}. As we shall discuss in detail, our work supersedes past evaluations. The ALP emissivity is made available in a public repository as a function of the ALP mass, temperature, and electron chemical potential of the plasma.

The paper is structured as follows. Section~\ref{sec:properties} reviews the properties of electrons and photons in a relativistic plasma. In Section~\ref{sec:emission} we argue that the pseudoscalar and derivative couplings are equivalent in the context of ALP-electron interactions. We identify for the first time that semi-Compton emission ($\gamma+e^\pm\rightarrow a+e^\pm$) is the dominant production channel across a wide range of ALP masses and plasma parameters, superseding bremsstrahlung and the spurious loop-induced Primakoff effect found in previous works. We also assess uncertainties in the production rate arising from electron thermal mass corrections, electron-electron Coulomb interactions, and the Landau-Pomeranchuk-Migdal effect. Section~\ref{sec:MFP} addresses the problem of trapping, which becomes relevant in the strong-coupling regime. Section~\ref{sec:results} presents our numerical results. 
Section~\ref{sec:constraints} is devoted to a collection of bounds on ALPs coupling to electrons (viz. from the observed cooling of SN~1987, from X-ray/gamma-ray flux of SN~1987A, from energy deposition, and from 511~keV line searches).
Finally, Section~\ref{sec:discussion} provides a discussion and summary of our findings.

\section{Properties of a dense plasma}
\label{sec:properties}
The plasma in the hypermassive remnants can reach temperatures of tens of MeV and electron chemical potentials of hundreds of MeV. Under these conditions, the dispersion relations of electrons and photons are significantly modified by the dense medium, directly affecting the emissivity of novel particles. In this section, we briefly summarize key results from historical studies on the main medium-induced modifications.

\subsection{Electrons}\label{sec:electrons}

The electron field is directly affected by the forward Compton scattering of medium photons. Because of this renormalization, two independent branches of excitations can be identified in the electron field, as discussed in detail by Braaten in Ref.~\cite{Braaten:1991hg} (see references therein for a historical overview): a collective plasmino, and the standard electron excitation. The parameter characterizing the medium effects for both of them is what Braaten calls the effective mass
\begin{equation}
    m_{\rm eff}=\frac{m_e}{2}+\sqrt{\frac{m_e^2}{4}+\frac{e^2(\mu_e^2+\pi^2 T^2)}{8\pi^2}}.
\end{equation}
We use here $\mu_e$ to denote the electron chemical potential; for a given electron number density $n_e=Y_e \rho/m_N$, where $Y_e$ is the electron fraction, $\rho$ is the mass density, and $m_N$ is the nucleon mass, we can extract the chemical potential from the definition of the number density
\begin{equation}
    n_e=\frac{Y_e \rho}{m_N}=-\frac{2T^3}{\pi^2}L_3\left(-e^{\mu_e/T}\right),
\end{equation}
where $L_3(x)$ is the polylogarithm of the third order. We have already assumed here electrons to be ultra-relativistic. In the limit $\mu_e\gg T$, this equation reduces to the conventional relation between density and chemical potential in a degenerate Fermi gas
\begin{equation}
    \mu_e=(3\pi^2 n_e)^{1/3}=\left(\frac{3\pi^2 Y_e \rho}{m_N}\right)^{1/3}.
\end{equation}
We will usually be concerned with the limit ${\mu_e \gg T \gg m_e}$ and $e^2 \mu_e^2 \gg m_e^2$, so that a simpler approximation is
\begin{equation}
    m_{\rm eff}\simeq \frac{e \mu_e}{2\sqrt{2}\pi}.
\end{equation}

For electrons, the dispersion relation changes from $\omega\simeq m_{\rm eff}+k/3$ in the non-relativistic limit to $\omega^2 \simeq k^2+2m_{\rm eff}^2$ in the ultra-relativistic limit, which is the one we are usually interested in. A special discussion must be made for positrons, whose typical energies are of the order of $T$; since $T\ll \mu_e$, it is possible that in certain regions positrons happen to be non-relativistic, with $T\lesssim m_{\rm eff}$. For our purposes, however, positrons mainly contribute to coalescence processes $e^+ e^-\to X$, where $X$ is the new particle. In the regions where $e^+$ are non-relativistic, their number density is suppressed as $e^{-m_{\rm eff}/T}$, and therefore such regions contribute the least to the overall coalescence signal. Therefore, it will suffice to adopt the simple dispersion relation $\omega = \sqrt{k^2 + 2m_{\rm eff}^2}$, which always holds for electrons and in the most relevant regions for positrons. Therefore, we will consistently treat electrons and positrons as having a thermal mass $m_{\rm th} = \sqrt{2} m_{\rm eff}$. As discussed in Ref.~\cite{Braaten:1991hg}, the wavefunction renormalization for electrons and positrons transitions from $Z \to 1/2$ in the non-relativistic limit to $Z \to 1$ for ultra-relativistic particles. Since we exclusively consider the latter, we will consistently set $Z = 1$ throughout.

A trickier aspect, however, is the spinor structure of the particle density matrix (for external particles) or propagator (for internal particles). For electrons, the renormalized propagator can be written as
\begin{equation}
    G(p)=\frac{1}{(1+a)\slashed{p}+b \slashed{u}-m_e(1+c)},
\end{equation}
where $p^\mu$ is the four-momentum of the particle, $u^\mu$ is the four-velocity of the frame in which the medium is at rest ($u^\mu=(1,0,0,0)$ for a stationary medium), $m_e$ is the vacuum electron mass, and $a$, $b$, and $c$ come from the thermal self-energy of the electron; explicit expressions for these functions are given in Ref.~\cite{Petitgirard:1991mf}. We will everywhere assume that $m_e$ is the smallest scale, compared to the thermal mass of the particles and their typical energies, so the term $m_e(1+c)$ can be neglected. The propagator can then be rewritten as
\begin{equation}\label{eq:green}
    G(p)=\frac{(1+a)\slashed{p}+b\slashed{u}}{(1+a)^2(E^2-p^2)+b^2+2b E },
\end{equation}
where we consider a stationary medium, and we take $p^\mu=(E,\bp)$; with a slight abuse of notation, we denote by $p=|\bp|$. An interesting question is whether the functions $a$ and $b$ can become large enough to affect the numerator. This issue was examined in Ref.~\cite{Carenza:2021pcm} in the context of the connection between pseudoscalar and derivative ALP-electron couplings. However, as we will discuss in Sec.~\ref{sec:pseudoscalar}, this connection is entirely independent of medium properties. Despite this, Ref.~\cite{Carenza:2021pcm} concluded that the functions $a$ and $b$ are generally very small.\footnote{It is not clear, however, what these functions were compared against. Notably, the function $b$ is not dimensionless, yet it is treated as such in Ref.~\cite{Carenza:2021pcm}. Moreover, only the integrated values of these functions over the momentum $|\bp|$ were considered, even though such integrated quantities do not directly appear in any physical context.} As we will now show, this conclusion has some subtleties, which are already apparent from the following observation: if $a$ and $b$ were truly negligible, how could the electron acquire a sizable in-medium mass?

To address this question, we can use the explicit expressions for the functions $a$ and $b$ in the limit of negligible $m_e$. These are provided in Ref.~\cite{Petitgirard:1991mf} in the limit of $T\gg \mu_e$, but the main features of our analysis remain true also in the limit of $\mu_e \gg T$, as long as both are much larger than $m_e$. In this limit, the functions $a$ and $b$ are given by
\begin{equation}
    a=\frac{m_{\rm eff}^2}{p^2}\left[1-\frac{E}{2p}\log\left(\frac{E+p}{E-p}\right)\right],
\end{equation}
and
\begin{equation}
    b=\frac{m_{\rm eff}^2}{2p}\left[\left(\frac{E^2}{p^2}-1\right)\log\left(\frac{E+p}{E-p}\right)-\frac{2E}{p}\right].
\end{equation}

Let us first consider the limit of non-relativistic electrons and positrons, for which $p$ is very small. In this case, $a\simeq -m_{\rm eff}^2/2Ep$ and $b\simeq -2 m_{\rm eff}^2/3E$, and the dispersion relation, given by the requirement that the denominator of Eq.~\eqref{eq:green} vanishes, requires $E=m_{\rm eff}$. So for an on-shell, non-relativistic electron, the function $a\sim m_{\rm eff}/p$ is actually very large, and the function $b\sim E$ is also not negligible. However, we are interested in the opposite limit of ultra-relativistic particles, with $E, p\gg m_{\rm eff}$. If the particle is off-shell, $a\sim m_{\rm eff}^2/p^2\ll 1$ and $b\sim m_{\rm eff}^2/p\ll p$, and therefore the renormalization factors can always be neglected. However, when the particles can be nearly on-shell, such arguments must be made more careful in dealing with the denominator of Eq.~\eqref{eq:green}. The term $bE$ can become comparable with $E^2-p^2$ because the latter becomes very small near the light cone, even though $E\sim p\gg m_{\rm eff}$. Indeed, in this case, neglecting $a\ll 1$ and $b^2\ll 2bE$, and keeping only the most important term in $b\simeq -m_{\rm eff}^2 E/p^2$,
the dispersion relation from the denominator of Eq.~\eqref{eq:green} becomes
\begin{equation}
    E^2-p^2-\frac{2m_{\rm eff}^2 E^2}{p^2}=0;
\end{equation}
since $E\simeq p$, we immediately recover our previous result that the denominator should be $E^2-p^2-2m_{\rm eff}^2$. In the numerator of the propagator (or in the density matrix of on-shell electrons), the terms depending on $a$ and $b$ can be safely neglected in this regime. So to summarize the propagator of electrons can be taken to be
\begin{equation}
    G(p)=\frac{\slashed{p}}{E^2-p^2-2m_{\rm eff}^2}.
\end{equation}
This gives a justification for treating them as particles with an effective mass $m_{\rm th}=\sqrt{2}m_{\rm eff}$, but crucially we are only authorized to keep this effective mass in the denominator of propagators, \textit{not} in their numerator, or in the density matrix of external electrons in a scattering process. So the choice of using a spinor structure of the form $\slashed{p}+m_{\rm th}$, as often done~\cite{Carenza:2021pcm, Carenza:2021osu, Ferreira:2022xlw}, only gives a misleading impression of increasing the precision; in reality, if one wanted to keep corrections due to the thermal mass of the medium, it is now clear that the correct density matrix would rather be of the form $\slashed{p}+b\slashed{u}$. In the limit of ultrarelativistic particles, the latter term should rather be neglected.

The plasmino excitation has a different dispersion relation, with a minimum at a momentum different from $0$, and asymptotically tending to the light cone at large momenta as $\omega\simeq k+2 e^{-\frac{k^2}{2m_{\rm eff}^2}-1}$. However, their contribution to physical processes is suppressed by a wavefunction renormalization factor $Z(k)$ which at large momenta decreases as $Z(k)\propto e^{-\frac{k^2}{2m_{\rm eff}^2}}$. Thus, ultrarelativistic plasminos can always be neglected. By our previous argument, this means that plasmino electrons can always be neglected, while plasmino positrons might contribute in the regions where they are non-relativistic ($T\lesssim m_{\rm eff}$), but such regions are the ones contributing less to coalescence due to Boltzmann suppression. Therefore, we will neglect altogether plasminos in this work, although it should be emphasized that the reasons are subtler than what was appreciated in previous works~\cite{Carenza:2021pcm,Ferreira:2022xlw}.

\subsection{Photons}

The excitations of the electromagnetic field in a plasma are well known to be strongly affected by the medium; the quantitative way this happens is comprehensively reviewed in Ref.~\cite{Raffelt:1996wa}. The primary process affecting the electromagnetic dispersion relation is forward scattering of photons on the electrons in the plasma, which causes transverse photons to acquire a modified dispersion relation. In addition, a new mode of excitation appears with a longitudinal polarization, corresponding to a density modulation of the plasma, which is generally known as a longitudinal plasmon. 
For photons, similarly to electrons, the impact of the medium can be measured by a single parameter, which is usually called the plasma frequency; for a relativistic plasma, this is
\begin{equation}
    \omega_P^2=\frac{4\alpha}{3\pi}\left(\mu_e^2+\frac{\pi^2 T^2}{3}\right).
\end{equation}
 As in the case of electrons, we will focus on those regions where $T\gg \omega_P$, since otherwise the population of photons is exponentially suppressed and generally contributes less to the emissivity. In these regions, the photons have an ultra-relativistic dispersion relation; for the transverse modes, this reads $\omega^2\simeq k^2+3\omega_P^2/2$, while the wavefunction renormalization factor is usually very close to 1. Thus, we can treat them effectively as ultrarelativistic particles with an effective mass $m_\gamma^2=3\omega_P^2/2$. For longitudinal modes, the frequency becomes ever closer to the light cone, with a dispersion relation $\omega\simeq k\left(1+2 e^{-2-\frac{2k^2}{3\omega_P^2}}\right)$.  At the same time, the wavefunction renormalization factor for longitudinal photons increases as $Z_L\simeq 2k^2/3\omega_P^2$. However, because of Ward's identity, a longitudinal photon for large $k$ becomes ever more polarized along the direction of the four-momentum, and therefore the probability for any process involving an external longitudinal photon is suppressed by $\omega-k\propto e^{-\frac{2k^2}{3\omega_P^2}}$. Therefore, for very large $k/\omega_P$, the increase in $Z_L$ is not sufficient to compensate for the exponential decrease in the coupling of the mode to the Standard Model; for ultrarelativistic plasmons, we can neglect the longitudinal degree of freedom, which indeed will be our assumption in this work. Notice that this statement refers \textit{only} to ultrarelativistic plasmons moving asymptotically close to the light cone. Thus, e.g. the resonant production of dark photons from longitudinal plasmons does not contradict the above statements, since the plasmons that match the dark photon dispersion relation must be not too close to the light cone. In our case, for ALP-electron coupling, longitudinal plasmons with $\omega\gg \omega_{\rm pl}$ must necessarily decouple because of Ward's identity.

\section{Emission of novel particles}
\label{sec:emission}

We can now discuss the main processes of emission of new particles from the SM plasma.  We will consider
a model where an ALP couples to electrons only via a
pseudoscalar coupling $g_a$
\begin{equation}
    \mathcal{L}_{\rm a}=\mathcal{L}_{\rm SM}+\frac{1}{2}\partial_\mu a \partial^\mu a-\frac{1}{2}m_a^2 a^2 +g_a a\bar{e}\gamma_5 e;
\end{equation}

\subsection{Pseudoscalar vs derivative coupling}\label{sec:pseudoscalar}

Since the earlier days of axion models, it was recognized that for many purposes the choice of a pseudoscalar coupling between ALPs and fermions is essentially equivalent to a derivative one. A notable exception to this rule, in the hadronic sector, is the case of baryons coupling to more than one Goldstone boson. In our case we have no such complications, since we are only examining the coupling of the ALP to the leptonic sector. For this case, the equivalence between the two interactions is easily proved by a simple redefinition of fields, as also done in Ref.~\cite{Raffelt:1996wa}. We start with a Lagrangian in which the ALP $a$ and electron field $\psi$ are coupled as
\begin{equation}
    \mathcal{L}=i\bar{\psi}\slashed{\partial}\psi-m_e \bar{\psi} e^{-i\frac{g_{ae} a}{m_e}\gamma^5}\psi,
\end{equation}
which to first order in $g_{ae}$ induces a pseudoscalar coupling between electrons and ALPs with coupling constant $g_{ae}$. The phase term can be removed by a field redefinition $\psi'=e^{-i\frac{g_{ae} a}{2m_e}\gamma^5}\psi$; at the classical level, this removes the ALP field from the mass term, so that the pseudoscalar coupling disappears, but leads to the appearance in the Lagrangian of an additional term
\begin{equation}\label{eq:L_der}
    \mathcal{L}_{\rm der}=-\frac{g_{ae}}{2m_e}\partial_\mu a \bar{\psi}'\gamma^\mu \gamma^5 \psi'.
\end{equation}
This shows that the equivalence between pseudoscalar and derivative coupling holds at the level of the fundamental Lagrangian. Reference~\cite{Carenza:2021pcm} attempted to show that the equivalence is true in the medium by applying the Dirac equation to the medium-renormalized electron field, but the above argument makes it clear that this is unnecessary; since the medium only changes the state over which the Green's functions are averaged, it \textit{must} preserve the equivalence between the two couplings, which, we emphasize, is valid only to first order in $g_{ae}$. In fact, this is also evident from the fact that the relation between the pseudoscalar and the derivative term must always involve the vacuum mass $m_e$, as clear from Eq.~\eqref{eq:L_der}, and not the in-medium mass. This aspect, which was considered somewhat puzzling in Refs.~\cite{Carenza:2021pcm,Ferreira:2022xlw}, appears entirely natural in light of the above argument, which clarifies that the equivalence is rooted at a more fundamental level than the medium on which physical properties are averaged.

One subtlety in the above argument is that the field redefinition we have performed is straightforward only at the classical level. In a quantum theory, it is well known that axial symmetry is explicitly broken by the chiral anomaly, meaning that even for a constant ALP field, the field redefinition should introduce an additional term in the Lagrangian. This term arises from the non-invariance of the path integral measure and takes the form
\begin{equation}\label{eq:alp-photon-chiral}
    \mathcal{L}_{\rm a\gamma}=-\frac{\alpha g_{ae} }{8\pi m_e}\epsilon^{\mu \nu \rho \sigma} F_{\mu\nu} F_{\rho \sigma} a=\frac{\alpha g_{ae}}{\pi m_e}\mathbf{E}\cdot\mathbf{B}a,
\end{equation}
where $F^{\mu\nu}$ is the electromagnetic field tensor, with the spatial electric $\mathbf{E}$ and magnetic $\mathbf{B}$ fields.
Thus, in passing from the pseudoscalar to the derivative representation, an ALP-photon coupling emerges. At first, this may seem puzzling, as it does not explicitly appear in the pseudoscalar version of the theory. However, this apparent paradox has a straightforward resolution: the original theory with a pseudoscalar interaction already contains an effective one-loop coupling to photons of precisely the same form (see Eqs.~(8b) and~(9a) in Ref.~\cite{Caputo:2021rux}) in the limit $m_a \to 0$, whereas the theory with a derivative interaction does not. Therefore, the effective ALP-photon coupling remains identical in both formulations.\\
The difference between the pseudoscalar and the derivative case in the one-loop ALP-photon coupling was noted in Ref.~\cite{Caputo:2021rux}, though the connection to the chiral anomaly was not explicitly established. The argument above clarifies that this difference has a fundamental origin. Since electrons do not couple to any other Nambu-Goldstone boson, we have now made it clear that the pseudoscalar and derivative interactions are truly equivalent to first order in $g_{ae}$, provided the anomalous ALP-photon coupling is kept in the transformation. Hereafter, we will assume that the theory with derivative coupling does not have a tree-level ALP-photon coupling, so that the term in Eq.~\eqref{eq:alp-photon-chiral} is canceled by an opposite term; this allows us to isolate the effects due to the electron term alone. With this choice, we are effectively defining the ``ALP coupling to electrons'' such that, at least in the ultra-violet range, for large energies, there is no surviving anomalous contribution to the ALP-photon contribution. There can still be additional one-loop ALP-photon couplings proportional to the ALP mass~\cite{Caputo:2021rux}, but as we will discuss in Sec.~\ref{sec:loops} these contributions do depend on the medium properties and turn out to be negligible.

\subsection{Semi-Compton emission}
For ALPs, a first channel of emission is semi-Compton scattering, through the process $e\gamma\to e a$.
To determine the spectrum of the emitted ALPs, we must also characterize their energy distribution in the scattering or, equivalently, the angular distribution of the emitted ALPs in the center-of-mass frame. Here, we assume that the scattering occurs in the ultra-relativistic regime, where the center-of-mass energy is much larger than the masses of the particles. In fact, typical center-of-mass energies are on the order of $\sqrt{\mu_e T}$, since the electron has typical energy $\mu_e$ and the photon has typical energy $T$. The electron thermal mass is of the order of $\sqrt{\alpha} \mu_e$. Therefore, we are assuming that $T\gg \alpha \mu_e$ (notice that, in order to consider the photons as relativistic, we were already assuming the more stringent condition $T\gg \omega_{\rm pl}\sim \sqrt{\alpha}\mu_e$) and $\sqrt{\mu_e T}\gg m_a$. The validity of these conditions is again limited only to the regions which dominate the emission. If photons cannot be considered relativistic, the emissivity would be dominated anyway by bremsstrahlung, due to the Boltzmann-suppressed photon population. Similarly, if the ALP mass becomes comparable to the center-of-mass energy, the total cross section is kinematically suppressed, making electron-positron coalescence the dominant contribution. Therefore, we adopt these approximations throughout, with the understanding that they may fail in regions that are not relevant for the overall emission. Additionally, this approximation breaks down when the incoming photon and electron are nearly collinear, as the available center-of-mass energy for scattering becomes small. However, the rate of these scatterings is suppressed by the relative velocity between the particles. As we shall see, their contribution to the total result is negligible in the angular integration we perform.

To proceed, we write the emitted number of ALPs as
    \begin{widetext}
\begin{align}\label{eq:axion_emissivity_formal}
    \frac{dn_a}{dt dV}=&\int \frac{d^3\bp}{(2\pi)^3 2E_\bp} \frac{d^3\bq}{(2\pi)^3 2E_\bq} \frac{d^3\bk}{(2\pi)^3 2E_\bk} \frac{d^3\bl}{(2\pi)^3 2E_\bl}
    f_e(p) f_\gamma(q) [1-f_e(p+q-E_a)](2\pi)^4\delta^{(4)}(p^\mu+q^\mu-k^\mu-l^\mu)|\mathcal{M}|^2,
\end{align}
    \end{widetext}
where $\bp$ and $\bq$ are the momenta of the incoming electron and photon respectively, $\bk$ and $\bl$ are the momenta of the outgoing ALP and electron, and the squared matrix element is summed over all spins and polarizations. We denote by $p^\mu$, $q^\mu$, $k^\mu$, $l^\mu$ the four-vectors associated with each three-vector, and with $p=|\bp|$ and the same for $q$, $k$, and $l$. The distribution functions $f_e(p)$, $f_\gamma(q)$ are the phase-space distributions for the species (Fermi-Dirac for electrons with temperature $T$ and chemical potential $\mu_e$, Bose-Einstein for photons with temperature $T$); they are the occupation number per single spin or polarization state. The matrix element $|\mathcal{M}|^2$ is summed over spins and polarizations of all particles.

We work in the lab frame, where we can use the delta function to integrate out the momentum $\bl$ and the angle between $\bk$ and $\bp+\bq$. The latter is
\begin{equation}\label{eq:cos_theta_compton}
    \cos\theta_{\bk,\bp+\bq}=\frac{2E_\bk E_\bp+2(E_\bk-E_\bp) q+2pqX-m_a^2}{2k\sqrt{p^2+q^2+2pqX}},
\end{equation}
where $X$ is the cosine of the angle between $\bp$ and $\bq$, and the integration should be performed only in the region in which this expression satisfies $|\cos\theta_{\bk,\bp+\bq}|<1$; this imposes the kinematic constraint on the scattering.
\begin{widetext}
    In this way, we are led to
\begin{equation}\label{eq:Compton_production}
    \frac{dn_a}{dE_adtdV}=\int_0^{+\infty} p dp \int_0^{+\infty} q dq \int_{-1}^{+1} dX \int_0^{2\pi} d\phi\frac{f_e(p) f_\gamma(q) [1-f_e(p+q-E_a)]}{512\pi^6 |\bp+\bq|}|\mathcal{M}|^2,
\end{equation}
where the ALP energy is denoted explicitly by $E_a\equiv E_\bk$. Here $\phi$ is the azimuthal angle of the vector $\bk$ around the direction of $\bp+\bq$.

For the matrix element, we separate out explicitly the contributions coming from the two diagrams corresponding to $u$-channel and $s$-channel, and the interference between the two

\begin{equation}
    |\mathcal{M}|^2=4\pi \alpha g_{ae}^2\left[\mathcal{F}_u+\mathcal{F}_s+\mathcal{F}_{us}\right],
\end{equation}

where
\edit{
\begin{eqnarray}
    &&\mathcal{F}_u=\frac{-\mathrm{Tr}\left[\gamma^\mu(\slashed{p}-\slashed{k})\gamma^5\slashed{p}(-\gamma^5)(\slashed{p}-\slashed{k})\gamma_\mu \slashed{l}\right]}{(m_{\rm th}^2-u)^2}=\frac{8\left[2(p-k)\cdot p (p-k)\cdot l-p\cdot l u\right]}{(m_{\rm th}^2-u)^2},\nonumber\\ 
    &&\mathcal{F}_s=\frac{-\mathrm{Tr}\left[\gamma^5 (\slashed{p}+\slashed{q})\gamma^\mu \slashed{p}\gamma_\mu (\slashed{p}+\slashed{q})(-\gamma^5)\slashed{l}\right]}{(s-m_{\rm th}^2)^2}=\frac{8\left[2(p+q)\cdot p (p+q)\cdot l-s p\cdot l\right]}{(s-m_{\rm th}^2)^2}, \\ \nonumber
    &&\mathcal{F}_{us}=\frac{-2\mathrm{Tr}\left[\gamma^5(\slashed{p}+\slashed{q})\gamma^\mu \slashed{p}(-\gamma^5)(\slashed{p}-\slashed{k})\gamma_\mu \slashed{l}\right]}{(s-m_{\rm th}^2)(u-m_{\rm th}^2)}=
    \frac{32p\cdot (p-k) l\cdot (p+q)}{(s-m_{\rm th}^2)(m_{\rm th}^2-u)}.
\end{eqnarray}

Finally, we can express all the four-products in the numerator in terms of the kinematic invariants $s$, $u$, and $t=m_a^2-s-u$ (since in the numerator we need only account for the ALP mass), obtaining

\begin{eqnarray}\label{eq:matrix_invariants}
    &&\mathcal{F}_u=-\frac{4su}{(m_{\rm th}^2-u)^2}\simeq \frac{4s}{m_{\rm th}^2-u},\nonumber\\ 
    &&\mathcal{F}_s=-\frac{4su}{(s-m_{\rm th})^2}\simeq -\frac{4u}{s-m_{\rm th}^2},\\ \nonumber
    &&\mathcal{F}_{us}=-\frac{8 (s-m_a^2) (m_a^2-u)}{(s-m_{\rm th}^2)(m_{\rm th}^2-u)}.
\end{eqnarray}
}
The corrections due to $m_{\rm th}^2$ need only be kept in the $u$-channel denominator $m_{\rm th}^2-u$, since for massless electrons and ALPs $u$ could nearly vanish for near-backward scatterings in which the ALP carries most of the electron energy. Even though the phase space integral cures the divergence at $X\simeq 1$, to avoid dealing with small values of the denominator in the numerical integration we also keep $m_{\rm th}$ corrections in $s$ channel denominator. So we have
\begin{align}
    s-m_{\rm th}^2=&\,2q(E_\bp-pX),\\
    \nonumber  m_{\rm th}^2-u=&\,2E_\bk E_\bp-m_a^2
    \\
    &-\frac{2pk}{\sqrt{p^2+q^2+2pqX}}\left[(p+qX)\cos\theta_{\bk,\bp+\bq}+q\sqrt{1-X^2}\sin\theta_{\bk,\bp+\bq}\cos\phi\right].
\end{align}
\end{widetext}

In the numerators, $m_{\rm th}^2$ can be neglected. The integral over $\phi$ can now be performed explicitly; it is convenient to use the notation $m_{\rm th}^2-u=A-B \cos \phi$, with
\begin{align}
    A&=2E_\bk E_\bp-m_a^2-\frac{2pk(p+q X) \cos\theta_{\bk,\bp+\bq}}{\sqrt{p^2+q^2+2pqX}},\;
    \nonumber \\
    B&=\frac{2pkq\sqrt{1-X^2} \sin\theta_{\bk,\bp+\bq}}{\sqrt{p^2+q^2+2pqX}}.
\end{align}
Then the result reads
\begin{align}\label{eq:compton_numerical_emissivity}
    \frac{dn_a}{dE_a dt dV}&=\int_0^{+\infty}pdp\int_0^{+\infty}qdq \int_{-1}^{+1}dX \nonumber
    \\
    &\times\frac{f_e(p) f_\gamma(q) \left[1-f_e(p+q-E_a)\right]}{512 \pi^6 |\bp+\bq|}4\pi \alpha g_{ae}^2I,
\end{align}

with
\edit{
\begin{align}\label{eq:compton_I_integral}
    I=8\pi &\Bigg[\frac{s}{\sqrt{A^2-B^2}}+\frac{A}{s-m_{\rm th}^2}
    \nonumber \\
    &-2\left(\frac{m_a^2}{\sqrt{A^2-B^2}}+1\right)\left(1-\frac{m_a^2}{s-m_{\rm th}^2}\right)\Bigg].
\end{align}

For later reference, it is useful to report also the expression for $I$ in the massless limit for all the particles involved,
\begin{align}\label{eq:compton_I_integral}
    I=8\pi&\Bigg[\frac{\sqrt{p^2+q^2+2pqX}}{|p-k|}
    \nonumber \\
   &+\frac{p(p+q X)-k(p-q)}{p^2+q^2+2pq X}-2\Bigg].
\end{align}
}
The apparent divergence at $p=k$ is caused by the neglect of the electron thermal mass, which becomes not permissible when $|p-k|\sim m_{\rm th}$, regularizing the divergence.

These expressions provide the emissivity under the only assumptions that $m_{\rm th}^2\ll pq(1-X)$ and $m_\gamma\ll q$, so that the electrons and photons can be taken to be ultra-relativistic. For our numerical computations, we therefore use them. For qualitative understanding, we note here that the matrix element is strongly enhanced when $u\sim m_{\rm th}^2$, which corresponds to the ALP carrying an energy $E_a$ comparable with the initial electron energy $p\sim \mu_e$. Indeed, if the scattering happened in vacuum, it is well known that Compton scattering of a photon off an electron leads to the final photon carrying most of the electron energy. However, in a medium, there is an opposite tendency due to the Pauli exclusion principle, because of which the electron cannot give away an energy of order $\mu_e$, which would require it to fall deep within the Fermi sea. Thus, in a very degenerate regime, the final ALP energy is rather of the order of $E_a\sim T\ll \mu_e$. In this case, we also have $q\ll p$; assuming also $m_{\rm th}^2/pq\ll q/p\ll 1$, the denominators can be approximated as
\begin{equation}
    s-m_{\rm th}^2\simeq m_{\rm th}^2-u\simeq  2pq(1-X),
\end{equation}
and then
\edit{
\begin{equation}
    |\mathcal{M}|^2\simeq 8\pi \alpha g_{ae}^2\frac{2k^2+2kq(1+3X)-q^2(1-3X^2)}{p^2}
\end{equation}
to lowest order in the small ratios $k/p$ and $q/p$. In this expression, we may replace $p$ with $p_F$, the Fermi momentum of the electrons, since the strongly degenerate distribution functions enforce electrons to lie at the Fermi surface. In turn, since the electrons are ultra-relativistic, we have $p_F\simeq \mu_e$. A subtle point here is that the denominator $p^{-2}$ causes a divergence, which is purely due to our assumption $p\gg k, q$. This means that the integral receives in reality also a contribution from the very small electron momenta, but such contribution would be suppressed by the degeneracy factor $1-f_e(p+q-E_a)\simeq e^{-\mu_e/T}$. The exponential suppression is much stronger than the power-law enhancement $(\mu_e/T)^{2}$, hence we may safely neglect this range.

We can now perform the integral over $p$ analytically; we also integrate over $X$, taking into account that the integral must be performed only over the region where $\cos\theta_{\bk,\bp+\bq}$ is in the physical region. From Eq.~\eqref{eq:cos_theta_compton}, we find that this requires $X$ to be between $X_{\rm min}$ and $1$, where in the massless limit
\begin{equation}\label{eq:kinematical_range_X}
    X_{\rm min}=\mathrm{max}\left[-1,1-\frac{2E_a(p+q-E_a)}{pq}\right].
\end{equation}
After performing the integration, we find

\begin{align}\label{eq:alp_compton_degenerate}
    \frac{dn_a}{dE_a dt dV}=\frac{\alpha g_{ae}^2}{8\pi^4\mu_e^2}\int_0^{+\infty}dq\frac{E_a q(E_a^2-q^2)}{(e^{q/T}-1)(e^{(E_a-q)/T}-1)}.
\end{align}

\begin{widetext}
    Thus, in the strongly degenerate limit, ALPs are emitted primarily around $E_a\sim T$. By performing the integral over $E_a dE_a$ numerically, we can now obtain the total energy emitted per unit volume and unit time
\begin{align}\label{eq:compton_strongly_degenerate}
    \frac{d\mathcal{E}_a}{dtdV}\simeq 3.7\times 10^{-3} g_{ae}^2 T^5\frac{T^2}{\mu_e^2}
   \simeq 1.1\times 10^{51}\;g_{ae}^2\; \left(\frac{T}{30\;\mathrm{MeV}}\right)^7\;\left(\frac{\mu_e}{150\;\mathrm{MeV}}\right)^{-2}\;\mathrm{erg}/\mathrm{cm}^3\;\mathrm{s}.
\end{align}
The regime of strong degeneracy corresponds to the contribution of electrons with $p\sim \mu_e$ interacting with photons $q\sim T\ll\mu_e$ and producing ALPs with $E_a\sim T$. In principle, ALPs with $E_a\sim T$ can also be produced by photons with momenta $q\sim \mu_e\gg T$. However, the contribution from this region of integration is ultimately subdominant. To see this explicitly, let us consider this region with $E_a\sim T\ll p,q\sim \mu_e$. In this case, the integral $I$ takes a very simple form
\begin{equation}
    I=\frac{8\pi q^2}{p(p+q)},
\end{equation}
and the range of values of $X$ is very narrow around $X=1$, with a width $\delta X\simeq 2 E_a (p+q)/pq$ (see Eq.~\eqref{eq:kinematical_range_X}). Thus, we can evaluate the function $I$ for $E_a=0$ and $X=1$, replace the integral over $dX$ with $\delta X$, and write the emissivity as    
\begin{equation}
    \frac{dn_a}{dE_a dt dV}=\frac{\alpha g_{ae}^2 E_a}{8\pi^4}\int_{E_a}^{+\infty}dp \int_{E_a}^{+\infty} dq \frac{1}{(e^{(p-\mu_e)/T}+1)(e^{(\mu_e-p-q)/T}+1)(e^{q/T}-1)}\frac{q^2}{p(p+q)}.
\end{equation}
\end{widetext}
The integral is cutoff at values of the order of $p,q\gtrsim E_a$, below which the approximations used are invalid, and is logarithmically sensitive to the lower cutoff. Due to the Bose-Einstein factor $(e^{q/T}-1)^{-1}$, the integral is exponentially suppressed if $q\sim \mu_e$, making the contribution from this range subdominant as anticipated.

Finally, for $E_a\sim \mu_e$, all of these approximations are of course invalid, since $E_a$, $p$, and $q$ are now all of the same order of magnitude. In this case, } when the ALP energy can become comparable with the electron energy, the integral $I$ is greatly enhanced when $p\sim k$. Therefore, keeping only the first term in square brackets of Eq.~\eqref{eq:compton_I_integral},
we may replace everywhere in the integrand $p=k$ except in the denominator; the integral $\int dp/|p-k|\sim 2\log(k/m_{\rm th})=\log(k^2/m_{\rm th}^2)$ since it must be cutoff at $|p-k|\sim m_{\rm th}$. For $k=p$ the integral over $X$ simply gives $2$ (see Eq.~\eqref{eq:kinematical_range_X}) and so we finally find
\begin{align}\label{eq:compton_mildly_degenerate}
    \frac{dn_a}{dE_a dt dV}=&\,\frac{\alpha g_{ae}^2}{8\pi^4}\log\left[\frac{E_a^2}{m_{\rm th}^2}\right]\frac{E_a}{e^{(E_a-\mu_e)/T}+1}\nonumber
    \\
  & \times \int_0^{+\infty}dq \frac{q}{(e^{q/T}-1)(1+e^{(\mu_e-q)/T})}.
\end{align}

If $\mu_e\gtrsim T$, the dominant contribution to the integral in Eq.~\eqref{eq:compton_mildly_degenerate} comes from $q\sim \mu_e$. We can then approximate the Bose-Einstein distribution with a Boltzmann one, and recognize the result as an integral over a Fermi-Dirac distribution for the electrons. For qualitative understanding, we can replace it with a strongly degenerate distribution, obtaining
\begin{equation}\label{eq:app_compton_first}
    \frac{dn_a}{dE_a dt dV}=\frac{\alpha g_{ae}^2\mu_e^2}{16\pi^4}e^{-\mu_e/T}\log\left[\frac{E_a^2}{m_{\rm th}^2}\right]\frac{E_a}{e^{(E_a-\mu_e)/T}+1}.
\end{equation}

We can also obtain an approximate expression for the total energy emitted $\mathcal{E}_a$ using Eq.~\eqref{eq:app_compton_first}
\begin{equation}\label{eq:compton_approximate}
    \frac{d\mathcal{E}_a}{dtdV}=\int dE_a E_a \frac{dn_a}{dE_a dt dV}=\frac{\alpha g_{ae}^2 \mu_e^5}{48\pi^4}\log\left[\frac{4\mu_e^2}{m_{\rm th}^2}\right]e^{-\mu_e/T}.
\end{equation}
As $\mu_e$ increases to $\mu_e\gg T$, the emissivity drops exponentially, until eventually it must be replaced by Eq.~\eqref{eq:alp_compton_degenerate}. We should also stress that since the photons that are dominating the emission have typical energies of order $q\sim \mu_e$, they are essentially guaranteed to respect our initial approximation $q\gg m_{\gamma}$. \\

Our analytical estimates provide a clear justification for neglecting semi-Compton emission from positrons. As we have seen, the emissivity is primarily driven by particles with typical energies of order $T$ in highly degenerate conditions, scaling as \edit{ $T^7/\mu_e^2$}, or by particles with energies of order $\mu_e$ in mildly degenerate conditions, scaling as $\mu_e^5 e^{-\mu_e/T}$. In the case of $e^+\gamma\to e^+ a$, however, the relevant particle energies remain of order $T$, as positrons follow a Maxwell-Boltzmann distribution. However, the number density of positrons is exponentially suppressed as $e^{-\mu_e/T}$. Thus, purely from dimensional analysis, the overall emissivity must scale as $T^5 e^{-\mu_e/T}$, making it generally negligible compared to the semi-Compton scattering on electrons considered here.

\subsection{Pair annihilation}
Electron-positron pairs can produce an ALP and a photon via the process $e^+ e^- \to a \gamma$. This channel becomes efficient at relatively high temperatures, where a sufficiently large population of positrons is present in the environment. Under such conditions, and provided the ALP mass is not too large---otherwise, coalescence processes would dominate the emission---the photon can be considered effectively massless, and the thermal mass of the electron remains much smaller than the COM energy. We treat the process of pair annihilation in direct analogy to the process of Compton scattering, adopting the same notation: $\bp$ for the initial electron momentum, $\bl$ for the initial positron momentum, $\bq$ for the photon momentum, and $\bk$ for the ALP momentum. Thus, 
\begin{widetext}
\begin{align}\label{eq:pair_emissivity_formal}
    \frac{dn_a}{dt dV}=\int \frac{d^3\bp}{(2\pi)^3 2E_\bp} \frac{d^3\bq}{(2\pi)^3 2E_\bq} \frac{d^3\bk}{(2\pi)^3 2E_\bk} \frac{d^3\bl}{(2\pi)^3 2E_\bl}
      f_e^{-}(p) f_{e^+}(q+E_a-p) [1+f_\gamma(q)](2\pi)^4\delta(p+l-k-q)|\mathcal{M}|^2.
\end{align}
\end{widetext}
Similarly to Compton, it is useful to integrate the positron momentum $\bl$ using the delta function, even if in this case this corresponds to an initial, rather than a final, particle. We then obtain
\begin{align}
    \frac{dn_a}{dE_a dt dV}&=\int pdpqdqdX d\phi 
    \nonumber
    \\
    &\times \frac{f_e(p) f_{e^+}(q+E_a-p)\left[1+f_\gamma(q)\right]}{512\pi^6|\bp-\bq|}|\mathcal{M}|^2,
\end{align}
where in this case $\phi$ is the azimuthal angle of $\bk$ around the direction of $\bp-\bq$.
We have once again used energy conservation to obtain the angle
\begin{equation}\label{eq:cosine_emission_pair}
        \cos\theta_{\bk,\bp-\bq}=\frac{2E_\bk E_\bp+2q(E_\bp-E_\bk)-2pqX-m_a^2}{2k\sqrt{p^2+q^2-2pqX}}
\end{equation}
(notice the symmetry with Eq.~\eqref{eq:cos_theta_compton}), and the integration over phase space is performed only over regions where $-1<\cos\theta_{\bk,\bp-\bq}<1$. In the massless limit, this condition translates in a range for $X$ between $X_{\rm min}$ and $1$, where
\begin{equation}\label{eq:minimum_X_pair}
    X_{\rm min}=\mathrm{max}\left[-1,1-\frac{2E_a(E_a+q-p)}{pq}\right].
\end{equation}

The squared matrix element, again summed over all spins and polarizations, \edit{can be obtained by straightforward calculations}
\begin{equation}
    |\mathcal{M}|^2=4\pi\alpha g_{ae}\left[\mathcal{F}_u+\mathcal{F}_t+\mathcal{F}_{ut}\right],
\end{equation}
with 
\begin{eqnarray}\label{eq:matrix_invariants_pair}
    &&\mathcal{F}_u\simeq \frac{-4t}{m_{\rm th}^2-u},\\ \nonumber
    &&\mathcal{F}_t\simeq\frac{-4u}{m_{\rm th}^2-t}\\ \nonumber
    &&\mathcal{F}_{ut}=\frac{8  (m_a^2-u) (m_a^2-t)}{(m_{\rm th}^2-t)(m_{\rm th}^2-u)}.
\end{eqnarray}
\edit{Note that, unlike in the Compton case, the interference term here contributes constructively, rather than destructively, to the emission.}
The values of $m_{\rm th}^2-u$ and $m_{\rm th}^2-t$ in the denominators must again contain the thermal mass corrections, so we have
\begin{widetext}
\begin{align}
    m_{\rm th}^2-t&=2q (E_\bp-pX),\\
    \nonumber  m_{\rm th}^2-u&=2E_\bk E_\bp-m_a^2
    -\frac{2pk}{\sqrt{p^2+q^2-2pq X}}\left[(p-qX)\cos\theta_{\bk,\bp+\bq}+q\sqrt{1-X^2}\sin\theta_{\bk,\bp+\bq}\cos\phi\right].
\end{align}
\end{widetext}

Similarly to the Compton case, the integral over $\phi$ can be performed explicitly
\begin{align}\label{eq:pair_numerical_emissivity}
    \frac{dn_a}{dE_a dt dV}=&\int_0^{+\infty}pdp\int_0^{+\infty}qdq \int_{-1}^{+1}dX \nonumber
    \\
   &\times \frac{f_e(p) f_{e^+}(q+E_a-p) \left[1+f_\gamma(q)\right]}{512 \pi^6 |\bp-\bq|}4\pi \alpha g_{ae}^2I,
\end{align}

with
\begin{align}\label{eq:pair_I_integral}
    I=\,&8\pi \Bigg[\frac{m_{\rm th}^2-t}{\sqrt{A^2-B^2}}+\frac{A}{m_{\rm th}^2-t}
    \nonumber
    \\
    &+2\left(\frac{m_a^2}{\sqrt{A^2-B^2}}+1\right)\left(1+\frac{m_a^2}{m_{\rm th}^2-t}\right)\Bigg],
\end{align}
where
\begin{align}
    A&=2E_\bk E_\bp-m_a^2-\frac{2pk(p-q X) \cos\theta_{\bk,\bp+\bq}}{\sqrt{p^2+q^2-2pqX}},\; 
    \nonumber\\
    B&=\frac{2pkq\sqrt{1-X^2} \sin\theta_{\bk,\bp+\bq}}{\sqrt{p^2+q^2-2pqX}}.
\end{align}

Also in this case it proves helpful to consider the integral $I$ in the massless limit
\begin{equation}
    I=8\pi\left[\frac{\sqrt{p^2+q^2-2pqX}}{|p-k|}+\frac{k(p+q)-p(p-qX)}{p^2+q^2-2pqX}+2\right].
\end{equation}

We use these equations for a numerical integration. When the ALP mass becomes comparable with the typical energies within the medium, the definition of the integral requires special attention. In fact, at sufficiently large ALP masses, the pair annihilation $e^+ e^-\to \gamma a$ can decompose in two reactions: $e^+\to e^+ \gamma$ and $e^+ e^-\to a$, which can both happen on-shell for special values of the parameters. In these cases, the denominators due to the electron propagators $u-m_{\rm th}^2$ and $t-m_{\rm th}^2$ can simultaneously vanish, leading to divergences within the integration range. A rigorous treatment would require a full thermal field-theoretical approach, renormalizing the electron propagator to account for medium correction. However, this is unnecessary, as the physical origin of the effect is clear: the divergence arises from the decomposition into two reactions, where the intermediate positron propagates on-shell. In the range of parameters in which this on-shell propagation is possible, one should not compute the ALP production from pair annihilation, but rather from the pure pair coalescence $e^+e^-\to a$ which is kinematically allowed and which we obtain in Sec.~\ref{sec:coalescence}. To resolve this issue, we simply need to determine the part of the integration range where both denominators can vanish, indicating that the reaction is kinematically allowed. The $t$-channel denominator vanishes when the photon can be Cherenkov-emitted: this happens either when it is emitted collinearly ($X=E_\bp/p$) or when it is soft ($q=0$). In the first case, from Eq.~\eqref{eq:cosine_emission_pair} we easily check that $|\cos\theta_{\bk,\bp-\bq}|>1$ always, and therefore the collinear divergence cannot show up in the integration range. In the second case, for $q=0$, we have
\begin{equation}
    \cos\theta_{\bk,\bp-\bq}=\frac{2E_\bk E_\bp-m_a^2}{2kp}.
\end{equation}
Since $\bq=0$, this is of course the angle at which the ALP must be emitted in order for the electron-positron coalescence to happen. 
We exclude the part of the integration range where $|\cos\theta_{\bk,\bp-\bq}|<1$, since here electron-positron coalescence is kinematically allowed and therefore should clearly be considered the relevant process, without accounting for pair annihilation.

For qualitative understanding, it is still useful to discuss the behavior in a degenerate environment with $T\lesssim \mu_e$. Here the $u$-channel denominator is smallest when $E_a\sim p$, so with ALP energies of order $\mu_e$. Instead, the $t$-channel denominator is smallest when $q\sim p$, and therefore the ALP has energies of order $T$ in this case. However, in this range $E_a\sim T\lesssim \mu_e$ the range of angles $X$ that is kinematically allowed from Eq.~\eqref{eq:minimum_X_pair} becomes narrow, of the order of $E_a/\mu_e$. Therefore, we will consider separately the contribution from the two ranges $E_a\sim \mu_e$ and $E_a\lesssim \mu_e$.

For $E_a\sim \mu_e$, we can perform the integral of the first term only in Eq.~\eqref{eq:pair_I_integral}, which is the dominant one when $E_a\simeq p$.  With logarithmic precision, $p$ can be replaced everywhere with $E_a$ in the integrand except in the denominator, where it produces the characteristic difference $p-E_a$ in the massless limit. After integration, this gives a logarithmic factor which must be cutoff at the upper limit at values of order $p-E_a\sim E_a$ and at the lower limit at values of order $p-E_a\sim m_{\rm th}$, where the mass cannot be neglected anymore. The range of $X$ for these values of $E_a\sim p$ is between $-1$ and $1$, so the integral is trivial, and we finally find
\begin{align}\label{eq:pair_spectrum_approximate}
    \frac{dn_a}{dE_a dt dV}=\,&\frac{\alpha g_{ae}^2}{8\pi^4}\log\left[\frac{E_a^2}{m_{\rm th}^2}\right]\frac{E_a}{e^{(E_a-\mu_e)/T}+1}
    \nonumber\\
    &   \times\int_0^{+\infty}dq \frac{q}{(e^{(q+\mu_e)/T}+1)(1-e^{-q/T})}.
\end{align}

The regime of $E_a\ll \mu_e$ is trickier than in the case of Compton production. The reason is that the $t$-channel contribution becomes divergent when $X\to 1$ and $p\to q$, corresponding to the photon being emitted collinear with the radiating electron. The divergence is cured by the neglected terms of order $m_{\rm th}^2$, which are of course included in the numerical calculations. Thus, obtaining an expansion for this regime is a complex task which ultimately may not be very instructive, as we obtain the emission numerically anyway, and pair annihilation turns out to be a subdominant process anyways. Thus, we do not consider the regime of $E_a\ll \mu_e$ further, and restrict ourselves to the regime of $E_a\sim \mu_e$.

Returning to Eq.~\eqref{eq:pair_spectrum_approximate}, we can approximately perform the integration in the degenerate limit noting that $e^{(q+\mu_e)/T}+1\sim e^{(q+\mu_e)/T}$ and obtaining
\begin{equation}\label{eq:approximate_pair_annihilation}
   \frac{dn_a}{dE_a dt dV}=\frac{\alpha g_{ae}^2 T^2}{48\pi^2}e^{-\mu_e/T}\log\left[\frac{4E_a^2}{m_{\rm th}^2}\right]\frac{E_a}{e^{(E_a-\mu_e)/T}+1}.
\end{equation}
In this case, the dominant contribution to the integral comes from $q\sim T$, so neglecting the photon mass requires $T\gg m_\gamma$. We can finally integrate over the ALP energies to obtain
\begin{equation}
   \frac{d\mathcal{E}_a}{dtdV}=\frac{\alpha g_{ae}^2 T^2 \mu_e^3}{144 \pi^2}\log\left[\frac{4\mu_e^2}{m_{\rm th}^2}\right]e^{-\mu_e/T}.
\end{equation}
By comparing with Eq.~\eqref{eq:compton_approximate}, we gather that Compton is generally expected to dominate over pair annihilation by a factor $\sim (\mu_e/T)^2$. We will later use these approximate expressions as a benchmark to validate our numerical integration.

\subsection{Electron-nucleus bremsstrahlung}

In the relatively colder regions of the star, both in NSM and SNe, the emission will be dominated by bremsstrahlung processes $eN\to eNa$. Notice that in these regions we can still take the electrons to be ultra-relativistic, since their typical energy is the chemical potential $\mu_e\gg m_{\rm th}\sim \sqrt{\alpha}\mu_e$. However, for bremsstrahlung we cannot make simple approximations for the typical energy of the emitted ALP. Therefore, we proceed to obtain the full expression for the emissivity. In this case, the COM frame does not offer particular advantages, since the nucleon is at rest in the laboratory frame, and its recoil energy can be neglected given its large mass, so it acts only as a sink of momentum. Therefore, we write the ALP emissivity directly in the laboratory frame, assuming the nucleon at rest. We assume an electron with momentum $\bp$ scattering with a nucleon and leaving with a final momentum $\bl$, after emitting an ALP with momentum $\bk$ and exchanging a momentum $\bq$ with the nucleon. We will denote by $E_\bp=\sqrt{|\bp|^2+m_{\rm th}^2}$, $E_\bl=\sqrt{|\bl|^2+m_{\rm th}^2}$, and $E_\bk=\sqrt{|\bk|^2+m_{a}^2}$ the energies of the electron (initial and final) and of the ALP.  The number of ALPs produced per unit time and volume is then
\begin{widetext}
    \begin{align}
    \frac{dn_a}{dtdV}=\int \frac{d^3\bp}{(2\pi)^32E_\bp}\frac{d^3\bl}{(2\pi)^32E_\bl}\frac{d^3\bk}{(2\pi)^32E_\bk}\frac{d^3\bq}{(2\pi)^3}
   |\mathcal{M}|^2(2\pi)^4\delta(\bp-\bq-\bl-\bk)\delta(E_\bp-E_\bl-E_\bk)f_{e^-}(E_\bp)[1-f_{e^-}(E_\bl)];
\end{align}

the squared matrix element $|\mathcal{M}|^2$ is summed over spins for all particles in the initial and final state. We can factorize the squared matrix element into a part relating to the electrostatic potential generated by the nuclei and the leptonic trace for the particle spins, so we find (we use the notation $q=|\bq|$, $p=|\bp|$, $l=|\bl|$)
\begin{align}
    |\mathcal{M}|^2=\frac{16\pi Z^2 \alpha^2 g^2 n_{\rm ion}}{q^2(q^2+k_D^2)} \times 4
    \bigg[\frac{\mathcal{T}_1}{(m_a^2-2E_\bp E_\bk+2\bp\cdot\bk)^2}&+\frac{\mathcal{T}_2}{(m_a^2+2E_\bl E_\bk-2\bl\cdot\bk)^2} \nonumber
    \\  &+\frac{2\mathcal{T}_3}{(m_a^2+2E_\bl E_\bk-2\bl\cdot\bk)(m_a^2-2E_\bp E_\bk+2\bp\cdot\bk)}\bigg].
\end{align}
\end{widetext}
The factor $4$ accounts for the sum over spins, so that the traces $\mathcal{T}_1$, $\mathcal{T}_2$, and $\mathcal{T}_3$ are only averaged over the spins; the factor $4\pi Z^2 \alpha/q^2(q^2+k_D^2)$ is the squared electrostatic potential generated by the nuclei, where $k_D^2=4\pi \alpha Z^2 n_{\rm ion}/T$ is the Debye screening scale. Here $n_{\rm ion}=n_e$ is the number density of ions; we assume for simplicity that the plasma is composed only of electrons and protons. Importantly, the screening is primarily produced by the less degenerate nucleons, not by the degenerate electrons. For this reason the screening prescription includes a $q^2 (q^2+k_D^2)$ denominator, rather than the $(q^2+k_D^2)^2$ often found in textbook treatments (see, e.g., Ref.~\cite{pitaevskii2012physical}); in passing through a single scattering event, an electron sees a fixed configuration of the much more slowly moving nuclei, and therefore only the squared electrostatic potential from the nuclei must be averaged (see the particularly clear discussion in Ref.~\cite{Raffelt:1996wa}). 

We will introduce for compactness the notation $D_l=m_a^2+2E_\bl E_\bk-2\bl\cdot\bk$ and $D_p=m_a^2-2E_\bp E_\bk+2\bp\cdot\bk$. The three terms in parenthesis correspond to the two Feynman diagrams and to their interference, and the Feynman traces are
\begin{eqnarray}
    &&\mathcal{T}_1=\frac{\mathrm{Tr}[\gamma^0(\slashed{p}-\slashed{k})\gamma^5\slashed{p}(-\gamma^5)(\slashed{p}-\slashed{k})\gamma^0 \slashed{l}]}{4}, \nonumber\\ 
    &&\mathcal{T}_2=\frac{\mathrm{Tr}[\gamma^5(\slashed{l}+\slashed{k})\gamma^0\slashed{p}\gamma^0(\slashed{l}+\slashed{k})(-\gamma^5)\slashed{l}]}{4},\\
    \nonumber && \mathcal{T}_3=\frac{\mathrm{Tr}[\gamma^5(\slashed{l}+\slashed{k})\gamma^0\slashed{p}(-\gamma^5)(\slashed{p}-\slashed{k})\gamma^0 \slashed{l}]}{4}.
\end{eqnarray}

We can perform the phase space integrations over $\bq$ and $l=|\bl|$ using the delta functions, to obtain the differential number of ALPs per unit energy $E_\bk$; however, the remaining angular integrations cannot all be performed trivially. The direction of $\bp$ can be integrated to give $4\pi$, and the azimuthal angle of $\bl$ around the direction of $\bp$ can also be trivially integrated to give $2\pi$. It remains to integrate over the angle between $\bk$ and $\bp$, which we denote by $x=\cos(\theta_{\bk\bp})$, and over the direction of the final electron; denoting the differential solid angle for the latter by $d\Omega_\bl$, we can finally obtain
\begin{align}\label{eq:axion_emission_bremsstrahlung}
    \frac{dn_a}{dE_a dtdV}=\int \frac{pkl Z^2 \alpha^2 g^2 n_{\rm ion}}{4\pi^4}f_{e^-}(E_\bp)[1-f_{e^-}(E_\bl)] dE_\bp
    \nonumber\\
    \times\int \frac{dx d\Omega_\bl}{q^2(q^2+k_D^2)}\left[\frac{\mathcal{T}_1}{D_p^2}+\frac{\mathcal{T}_2}{D_l^2}+\frac{2\mathcal{T}_3}{D_p D_l}\right].
\end{align}
The momentum of the external field $q$ is here implicitly determined by momentum conservation as
\begin{equation}
    q^2=p^2+l^2+k^2-2\bp\cdot\bk-2\bp\cdot\bl+2\bk\cdot\bl.
\end{equation}

\begin{widetext}
    The traces can be computed explicitly, and are most easily expressed by introducing the unit vectors $\bn_p$, $\bn_l$, and $\bn_k$ in the directions of $\bp$, $\bl$, and $\bk$, so that
\begin{eqnarray}
    && \mathcal{T}_1=-[lp(m_a^2-2E_\bk^2+2k E_\bk x)+\bn_p \cdot \bn_l m_a^2 l p+\bn_l\cdot\bn_k 2pl k (kx-E_\bk)],\\ \nonumber 
    && \mathcal{T}_2=[2pl E_\bk (E_\bk+kx)-pl m_a^2-\bn_l\cdot \bn_p m_a^2 l p-2\bn_l\cdot \bn_k p l k(E_\bk+k x)],\\ \nonumber
    && \mathcal{T}_3=-[m_a^2 pl+\bn_p\cdot\bn_l l p (m_a^2-2E_\bk^2)+\bn_l\cdot \bn_k 2lpk^2 x].
\end{eqnarray}
\end{widetext}
The reason for this apparently complex three-dimensional notation is that we can explicitly see now that all the integrals over $d\Omega_\bl$ involve integrand functions that are both in the numerator and denominator polynomials in the variables $\bn_l\cdot\bn_p$ and $\bn_l\cdot\bn_k$. Such integrals, although laboriously, can be performed entirely analytically, so that the emissivity can be reduced to an integral over $E_\bp$ and $x$ only
\begin{align}
    \frac{dn_a}{dE_a dtdV}=\int \frac{pkl Z^2 \alpha^2 g^2 n_{\rm ion}}{4\pi^4}f_{e^-}(E_\bp)[1-f_{e^-}(E_\bl)] dE_\bp
    \nonumber\\
    \int dx[\mathcal{I}_1+\mathcal{I}_2+2\mathcal{I}_3].
\end{align}

The complete expressions for the functions $\mathcal{I}_1$, $\mathcal{I}_2$, and $\mathcal{I}_3$ are given in the appendix. We use these for our numerical evaluations. On the other hand, in the limit of massless ALPs, it is still useful to obtain approximate expressions to understand the qualitative dependence of bremsstrahlung rates on the environment. Let us consider two extreme cases; that of strong screening $k_D\gg m_{\rm th}$ and of weak screening $k_D\ll m_{\rm th}$. In the end we will be interested in a somewhat intermediate regime, with $k_D\gtrsim m_{\rm th}$. However, it is important to understand the angular distribution of the emitted radiation in both cases, especially because of our later considerations on the Landau-Pomeranchuk-Migdal suppression for this process.

\begin{figure*}
	\centering
	\includegraphics[width=1\textwidth]{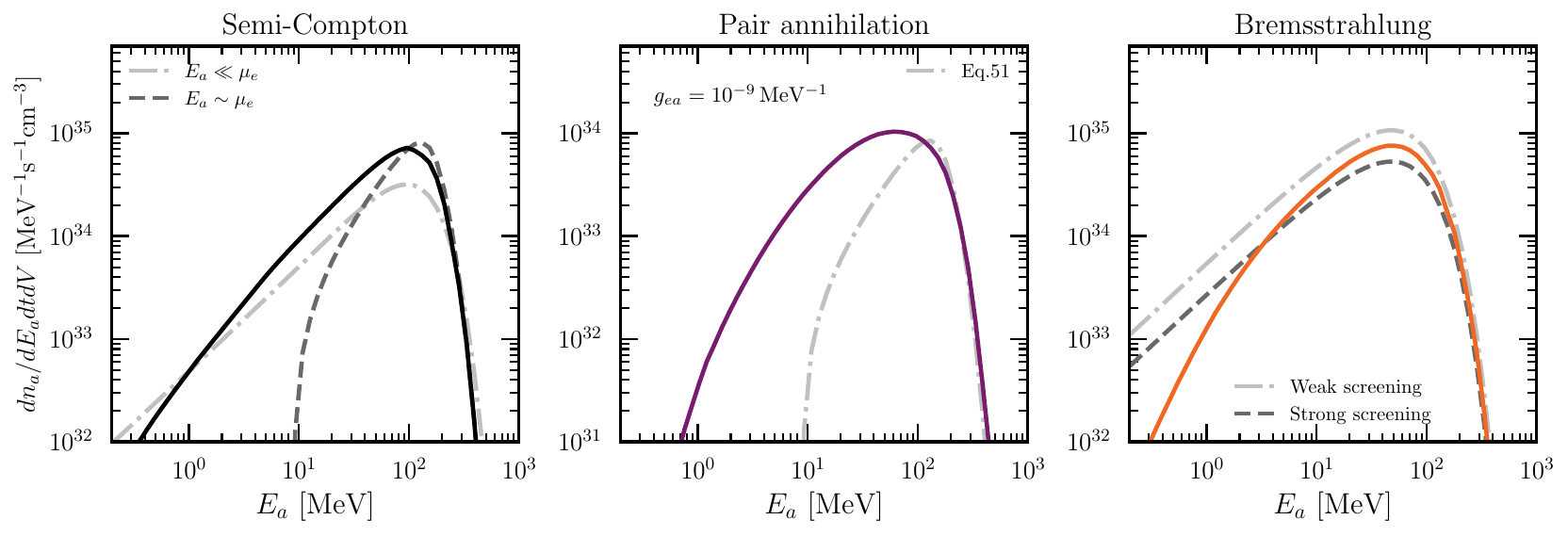}
	\caption{ALP emissivity for $m_{a}=100$~keV, $T=30\,{\rm MeV}$, $\mu_e=150\,{\rm MeV}$. We separately show the emissivity for all different processes; numerical results are shown as solid lines. We compare with the analytical approximations: for semi-Compton, with the high-$E_a$ (Eq.~\eqref{eq:app_compton_first}) and low-$E_a$ (Eq.~\eqref{eq:alp_compton_degenerate}); for pair annihilation, with Eq.~\eqref{eq:approximate_pair_annihilation}; for bremsstrahlung, with the weak screening (Eq.~\eqref{eq:spectrum_bremsstrahlung}) and the strong screening (Eq.~\eqref{eq:strong_screening_brem_app}) approximations.}
	\label{Fig: All_emissivities_ma=100KeV}
\end{figure*}

\subsubsection{Weak screening: $k_D\ll m_{\rm th}$}

We start with the angular integral
\begin{equation}
    I=\int \frac{dxd\Omega_\bl}{q^2(q^2+k_D^2)}\left[\frac{\mathcal{T}_1}{D_p^2}+\frac{\mathcal{T}_2}{D_l^2}+\frac{2\mathcal{T}_3}{D_p D_l}\right].
\end{equation}
We start with the physical observation that in the limit of very small masses the angles that $\bp$ and $\bl$ form with $\bk$, which we shall call $\theta_\bp$ and $\theta_\bl$, are very small, so that the denominators $D_p$ and $D_l$ are both very small. Thus, we define $\delta_\bp=\theta_\bp p/m_{\rm th}$ and $\delta_\bl=\theta_\bl l/m_{\rm th}$ and expand the integrand for very small $m_{\rm th}^2$; as usual, in the numerators $\mathcal{T}_1$, $\mathcal{T}_2$, and $\mathcal{T}_3$, we must neglect all higher orders in the electron thermal mass, while in the denominators we find $D_p\simeq -m_{\rm th}^2 k(1+\delta_p^2)/p$, and $D_l\simeq m_{\rm th}^2 k (1+\delta_l^2)/l$. The momentum of the external field in this expansion can be written
\begin{equation}
    \frac{q^2}{m_{\rm th}^2}=\delta_p^2+\delta_l^2-2\delta_p \delta_l\cos\phi+ \xi^2,
\end{equation}
with $\phi$ the azimuthal angle between the plane of $\bk$ and $\bp$ and the plane of $\bk$ and $\bl$, and
\begin{equation}
    \xi=m_{\rm th}\left[\frac{1+\delta_l^2}{2l}-\frac{1+\delta_p^2}{2p}\right].
\end{equation}
\begin{widetext}
    Therefore, the entire integral can be rewritten as
\begin{equation}\label{eq:approximate_integral}
    I=\frac{2}{m_{\rm th}^2 p l}\int \frac{\delta_p d\delta_p \delta_l d\delta_l d\phi}{\frac{q^2}{m_{\rm th}^2}\left[\frac{q^2}{m^2_{\rm th}}+\lambda^2\right]} \left[\frac{\delta_p^2}{(1+\delta_p^2)^2}+\frac{\delta_l^2}{(1+\delta_l^2)^2}-\frac{2\delta_p \delta_l \cos\phi}{(1+\delta_p^2)(1+\delta_l^2)}\right],
\end{equation}
with $\lambda=k_D/m_{\rm th}$. The denominator $q^2$ becomes very small when $\delta_l\simeq \delta_p$ and $\phi\simeq 0$, which is the region that dominates the integration. Therefore, we will expand for very small $\Delta=\delta_l-\delta_p$ and $\phi$, keeping only the leading terms in the expansion; we will see that this corresponds to keeping only the dominant logarithmic terms in the final result. Thus we find
\begin{equation}
    I=\frac{2}{m_{\rm th}^2 p l}\int d\delta_p d\delta_l d\phi \frac{1}{\delta_p^2\left[\phi^2+\frac{\xi^2+\lambda^2+\Delta^2}{\delta_p^2}\right]\left[\phi^2+\frac{\xi^2+\Delta^2}{\delta_p^2}\right]}\frac{(1-\delta_p^2)^2\Delta^2+\delta_p^2 \phi^2}{(1+\delta_p^2)^4}.
\end{equation}
We can now perform the integral over $\phi$ first, obtaining
\begin{equation}
    I=\frac{\pi}{m_{\rm th}^2 p l}\int \frac{d\delta_p d\Delta \delta_p}{(1+\delta_p^2)^4 (\sqrt{\Delta^2+\xi^2}+\sqrt{\Delta^2+\xi^2+\lambda^2})}\left[\frac{(1-\delta_p^2)^2 \Delta^2}{ \sqrt{\Delta^2+\xi^2}\sqrt{\Delta^2+\xi^2+\lambda^2}}+1\right].
\end{equation}
\end{widetext}

The integral over $\Delta$ now diverges at $\Delta$ large, but this is only because we have already expanded for very small $\Delta$, and the original integral in Eq.~\eqref{eq:approximate_integral} has no divergence. We see now that throughout most of the range the integral over $\Delta$ is logarithmic, and therefore the divergence at large $|\Delta|$ only affects the upper limit of the integration; we can integrate up to $|\Delta|\sim 1$ within logarithmic precision. At small $|\Delta|$, with logarithmic precision we can simply integrate down to $|\Delta|\sim \mathrm{max}(\xi,\lambda)$, where $\xi$ can be evaluated for vanishing $\delta_p$ and $\delta_l$ as $\xi\simeq m_{\rm th}k/2pl$. Therefore, the integral over $\Delta$ simply gives
\begin{equation}
    I\simeq \frac{2\pi L}{m_{\rm th}^2 p l}\int d\delta_p \delta_p \left[\frac{(1-\delta_p^2)^2+1}{(1+\delta_p^2)^4}\right],
\end{equation}
where $L=\log(1/\mathrm{max}(\xi,\lambda))$; the factor $2$ appears because of the integration for both positive and negative $\Delta$.

The integral over $\delta_p$ is now trivial and we finally get
\begin{equation}
    I\simeq \frac{2\pi L}{3 m_{\rm th}^2 p l}.
\end{equation}
This expression is obtained in the leading logarithmic expansion $L\gg 1$, which is quite unjustified in this context because usually $\lambda\sim 10$, so it can only serve as a qualitative guidance.

The angular distribution for the radiation is particularly interesting. In the regime of weak screening, both $\delta_p$ and $\delta_l$ are of order $1$, and in fact the emission is dominated by the region where they are very close together as we have seen. The physics behind this conclusion is very clear: the relativistic radiation is strongly beamed, and because the scattering is weakly screened, it is dominated by the region of large impact parameters and small deflection angles, so the radiation can remain at small angles with the electron both before and after the scattering.

\sloppy After we replace this expression for $I$ in Eq.~\eqref{eq:axion_emission_bremsstrahlung}, we find
\begin{align}\label{eq:spectrum_bremsstrahlung}
    \frac{dn_a}{dE_a dt dV}=&\,\frac{Z^2 \alpha^2 g^2_{ae}n_{\rm ion} E_a}{6\pi^3 m_{\rm th}^2}
    \nonumber\\
   &    \times \int_0^{+\infty}dp f_{e^-}(p) \left[1-f_{e^-}(p-E_a)\right]L.
\end{align}
We see that in the soft limit $E_a\ll p$ the ALP spectrum has a characteristic form ${dn_a/dE_a dt dV\propto E_a}$; this feature is easily interpreted in comparison with photon bremsstrahlung, for which instead $dn_a/dE_a dt dV\propto E_a^{-1}$, because the squared matrix element is suppressed by a factor $(E_a/p)^2$. In addition, the typical interaction rate is not suppressed as $p^{-2}$ at large energies, as for Klein-Nishina scattering, but rather is momentum-independent and characterized by the factor $m_{\rm th}^{-2}$, something that we also noticed in Refs.~\cite{Diamond:2023cto,Diamond:2023scc} in the context of electron-electron bremsstrahlung.
The logarithm $L$ can be evaluated for $p\sim l \sim \mu_e$, since its argument was anyway obtained in order of magnitude only, and the resulting integral over $p$ can then be done explicitly to give
\begin{equation}
    \frac{dn_a}{dE_a dt dV}=\frac{Z^2 \alpha^2 g^2_{ae}n_{\rm ion} E_a T L}{6\pi^3 m_{\rm th}^2 (e^{E_a/T}-1)}\log\left[\frac{e^{(\mu_e+E_a)/T}+1}{e^{\mu_e/T}+1}\right].
\end{equation}
We can now integrate to determine the total energy emitted by bremsstrahlung; in doing so, we notice that the majority of ALPs are emitted for $E_a\sim T$, so if $T\lesssim \mu_e$ we may approximate the last square bracket as $E_a/T$ (this is a very good approximation for any value of $E_a$), finally obtaining
\begin{equation}
    \frac{d\mathcal{E}_a}{dtdV}=\frac{Z^2\alpha^2 g_{ae}^2 \pi n_{\rm ion}T^4 L}{90m_{\rm th}^2}.
\end{equation}

\begin{figure*}
	\centering
	\includegraphics[width=0.5\textwidth]{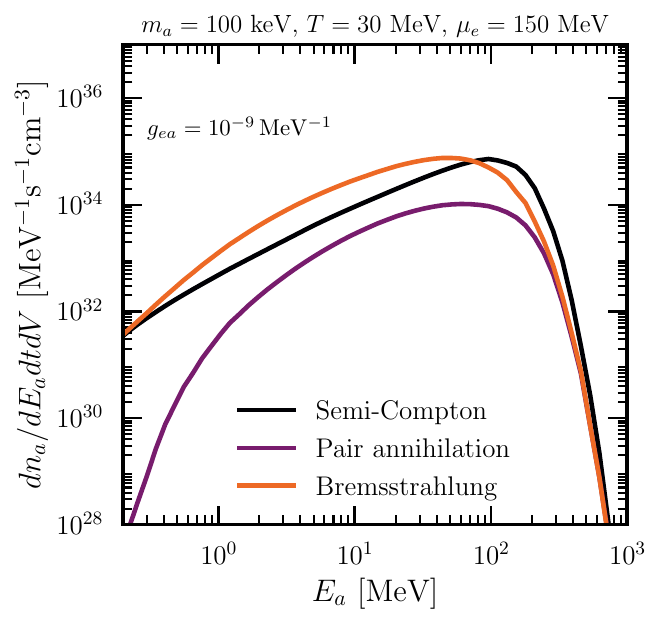}
	\caption{ALP emissivity for $m_{a}=100$~keV, $T=30\,{\rm MeV}$, $\mu_e=150\,{\rm MeV}$, separated by components. The coalescence contribution is zero since  $m_a < 2\,m_{\rm th}$.}
	\label{Fig:All_emissivities_together}
\end{figure*}

\subsubsection{Strong screening: $k_D\gg m_{\rm th}$}

In the opposite limit of strong screening, we consider again the original integral Eq.~\eqref{eq:approximate_integral}, where now $\lambda\gg 1$ is the dominant scale in the denominator. This means that the logarithmic divergence when $\delta_p\simeq \delta_l$ is now eliminated. Physically, the scattering is so strongly screened that the deflection angle between the initial and final electron is dominated by the regions of order $\lambda\gg 1$. Therefore, the radiation cannot remain at small angles with both the initial and the final electron. Nevertheless, the emission is still dominated by regions where the radiation is at small angles with either $\bp$ and $\bl$; we can picture therefore the emission as strongly beamed in the direction of the radiating electron, either the initial or the final one. Motivated by this picture, let us consider $\delta_p\sim 1$ and perform the integration over $\delta_l$ not restricted to be of order $1$. We have to separate the integration in two separate regions, with $\delta_l<\delta_0$ and $\delta_l>\delta_0$. We choose $1\ll \delta_0\ll \lambda$, so that the relevant approximations can be performed in each separate regime. Therefore we write
\begin{equation}
    I=\frac{2}{m_{\rm th}^2 p l}\int \delta_p d\delta_p\left[J_1+J_2\right],
\end{equation}
where $J_1$ and $J_2$ are the integrals performed over the first and the second region respectively.
\begin{widetext}
In the first integration region, we have $q^2/m_{\rm th}^2\ll \lambda^2$, so we can approximate
\begin{equation}
    J_1=\int_0^{\delta_0}\delta_l d\delta_l \int_0^{2\pi} d\phi\frac{m_{\rm th}^2}{\lambda^2 q^2}\left[\frac{\delta_p^2}{(1+\delta_p^2)^2}+\frac{\delta_l^2}{(1+\delta_l^2)^2}-\frac{2\delta_p \delta_l \cos\phi}{(1+\delta_p^2)(1+\delta_l^2)}\right].
\end{equation}
There is no logarithmic enhancement to this integral when $\delta_p\sim \delta_l$ or $\phi\sim 0$, so the integral over $\phi$ must now be performed exactly. We then find
\begin{equation}
    J_1=\int_0^{\delta_0}\frac{d\delta_l}{2\lambda^2 \delta_p}\frac{4\pi\delta_l \delta_p}{(1+\delta_l^2)^2(1+\delta_p^2)^2}\left[1+\delta_p^2+\delta_l^2+\delta_p^2 \delta_l^2-\delta_p\delta_l \sqrt{\frac{\delta_l^2}{\delta_p^2}+\frac{\delta_p^2}{\delta_l^2}-2}\right].
\end{equation}
\end{widetext}
This integral is logarithmic at $\delta_0\gg 1$. The integration can be done analytically and obtained in the limit $\delta_0\gg 1$
\begin{equation}
    J_1=\frac{1}{2\lambda^2 \delta_p}\left[\frac{4\pi\delta_p^3}{(1+\delta_p^2)^2}\log\delta_0+\frac{2\pi\delta_p}{1+\delta_p^2}\right].
\end{equation}
In the integral $J_2$, we cannot neglect $q^2/m_{\rm th}^2$ compared to $\lambda^2$, since $\delta_l\gg 1$; however, since we are interested in the angular distribution for $\delta_p\sim 1$, we can neglect everywhere $\delta_p$ compared to $\delta_l$. In the squared parenthesis of Eq.~\eqref{eq:approximate_integral}, only the first term survives, since the others are smaller. This corresponds to the radiation being emitted only by the initial electron with which it is nearly collinear, while the final electron is deflected outside of the radiation cone. Under these approximations, $\phi$ disappears from the integrand, and therefore the integral over $\phi$ simply gives $2\pi$. So we have
\begin{equation}
    J_2=\int_{\delta_0}^{+\infty}\frac{\delta_l d\delta_l}{\delta_l^2 (\delta_l^2+\lambda^2)}2\pi \frac{\delta_p^2}{(1+\delta_p^2)^2}.
\end{equation}
The integral is trivial, and can be performed in the limit $\lambda\gg \delta_0$ giving
\begin{equation}
    J_2=\frac{2\pi \delta_p^2}{(1+\delta_p^2)^2\lambda^2}\log\left(\frac{\lambda}{\delta_0}\right).
\end{equation}
We can now sum the contribution  between $J_1$ and $J_2$. The dependence on the arbitrary $\delta_0$ drops out, as it should. We can keep only the logarithmically enhanced term to find
\begin{equation}
    I=\frac{4\pi\log\lambda}{m_{\rm th}^2 p l \lambda^2}\int d\delta_p\frac{\delta_p^3}{(1+\delta_p^2)^2}.
\end{equation}
The integral is logarithmically divergent at infinity, because all of our approximations only hold for $\delta_p\ll \lambda$. Therefore, we may limit our integration only up to values of order $\delta_p \sim \lambda$ (one could of course be more precise and explicitly use different approximations for a regime $\delta_p\sim \lambda$, but the logarithmically enhanced terms can be gathered by this simpler method) and finally obtain
\begin{equation}
    I=\frac{4\pi \log^2(k_D/m_{\rm th})}{k_D^2 p l}.
\end{equation}

Thus, the emitted ALP spectrum changes from the previous case as
\begin{align}\label{eq:strong_screening_brem_app}
    \frac{dn_a}{dE_a dt dV}=\frac{Z^2 \alpha^2 g^2_{ae}n_{\rm ion} E_a T \log^2(k_D/m_{\rm th})}{\pi^3 k_D^2 (e^{E_a/T}-1)}
    \nonumber\\
    \times\log\Bigg[\frac{e^{(\mu_e+E_a)/T}+1}{e^{\mu_e/T}+1}\Bigg].
\end{align}
Similarly, the total energy emitted in ALPs is
\begin{equation}\label{eq:bremsstrahlung_emissivity_strong_screening}
    \frac{d\mathcal{E}_a}{dtdV}=\frac{Z^2\alpha^2 g_{ae}^2 \pi n_{\rm ion}T^4 \log^2(k_D/m_{\rm th})}{15k_D^2}.
\end{equation}
So the main change in the strong screening limit is that the cross section is not anymore of the order of $\alpha^2/m_{\rm th}^2$, but rather $\alpha^2/k_D^2$. The reason is that the minimum angle at which an ALP can be radiated away is not anymore determined by the mass of the electron during the radiation phase after the collision, but rather by the scattering angle during the collision. This leads to a general suppression of the bremsstrahlung cross section.

In a SN environment, the Debye screening scale is of order $k_D^2\sim \alpha \mu_e^3/T$, compared to the thermal mass $m_{\rm th}^2\sim \alpha \mu_e^2$, so we generally expect $\lambda \gtrsim 1$. Clearly our approximations here cannot be taken too seriously in this intermediate regime, especially because the logarithm $\log\lambda$ is not at all large, and therefore is not a dominant term; yet we can use these equations to infer the order of magnitude of the resulting emission and its scaling with the medium parameters. From Eq.~\eqref{eq:bremsstrahlung_emissivity_strong_screening} we see that the radiated energy per unit volume turns out to be of the order of $d\mathcal{E}_a/dtdV\sim \alpha g_{ae}^2 T^5$. From Eq.~\eqref{eq:compton_strongly_degenerate}, we gather that also the Compton emissivity scales in proportion to the same parameters, so that even in highly degenerate environments Compton and bremsstrahlung emission can compete. In fact, bremsstrahlung emission suffers an additional form of suppression, due to nucleon degeneracy---which of course does not affect Compton, in which there is only one target that is degenerate, the electron one---which we now discuss. Overall, our numerical results show that even at very large chemical potentials bremsstrahlung never turns out to completely dominate the emissivity.

\begin{figure*}
	\centering
	\includegraphics[width=1\textwidth]{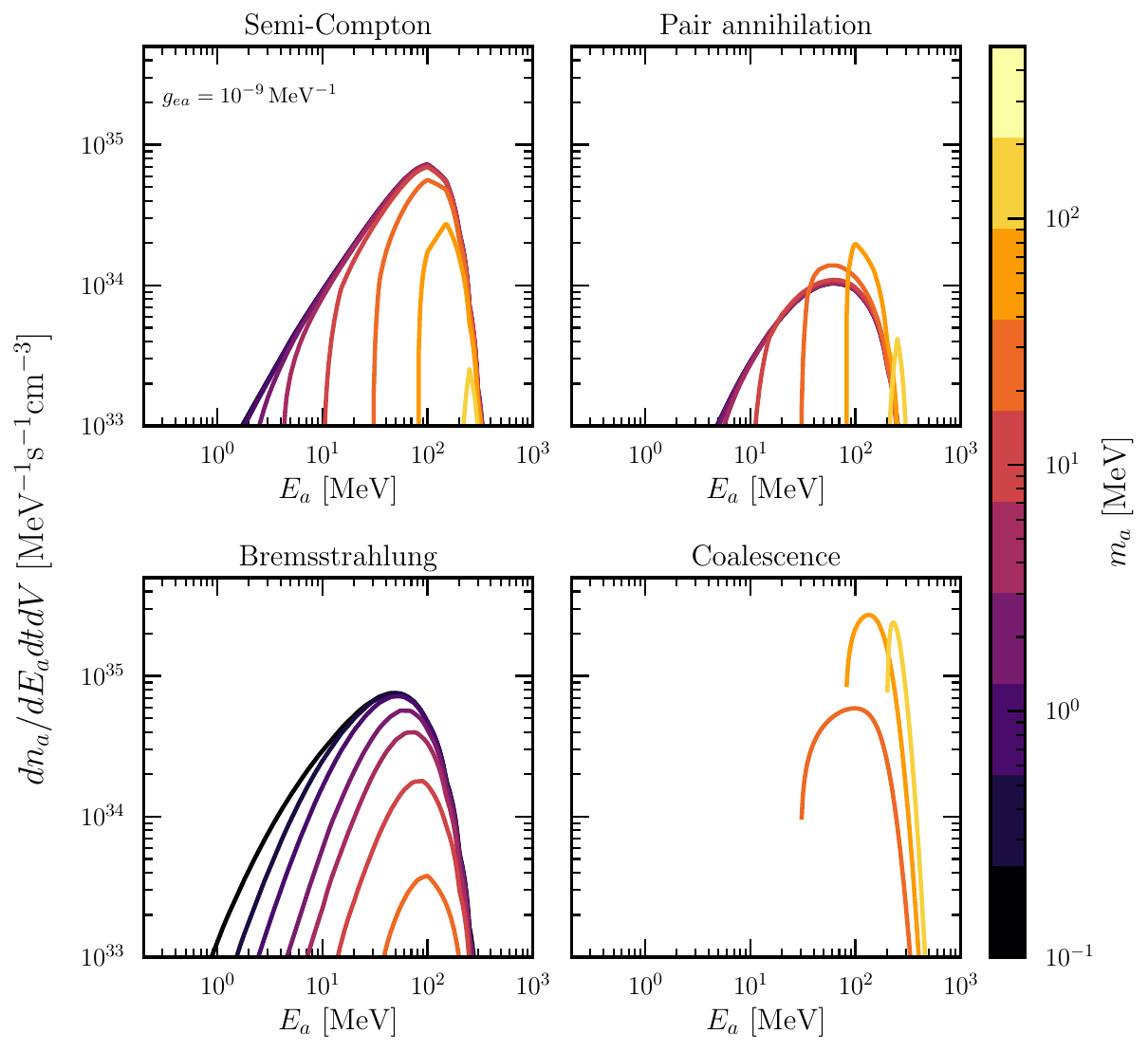}
	\caption{Emissivities for $T=30\,{\rm MeV}$, $\mu_e=150\,{\rm MeV}$, and for varying ALP mass.}
	\label{Fig:All_emissivities_different_masses}
\end{figure*}

\subsubsection{Nucleon degeneracy}

Throughout our previous calculations, we have assumed a non-degenerate proton population with $n_{\rm ion} = n_e$. However, when $\mu_e$ becomes sufficiently large, nucleon degeneracy can introduce significant corrections. To account for this effect, and following our general assumption that nucleons are infinitely massive and thus exchange negligible energy in the bremsstrahlung process, we introduce the suppression factor
\begin{equation}
    S=\frac{\int f_p(E) \left[1-f_p(E)\right]E^2 dE}{\int f_p(E) E^2 dE},
\end{equation}
where $f_p(E)$ is the proton phase space distribution as a function of energy.

\subsection{Main uncertainties in the production rates of ALPs}

Let us briefly comment on the main approximations in our calculation to identify the primary sources of uncertainty.

All the processes discussed so far are characterized by an electron radiating an ALP while moving with ultra-relativistic velocities. For pair annihilation, the ingoing positron can be regarded as an electron moving backward in time, so the following arguments still apply.  Landau and Pomeranchuk~\cite{landau1953limits} pointed out that such scenarios require special consideration, as the typical length scales over which the electron radiates are significantly elongated due to the near-collinearity between the electron and the radiated field. Essentially, since the emitted particle remains nearly collinear with the electron, it takes a long time to clearly separate from it.
As a result, treating each scattering event as independent within a dense medium requires stricter conditions than usual. We now discuss these conditions for the processes considered in this work.

The formation length of an ALP in vacuum is defined as the length over which the phase difference accumulated by the ALP after its emission becomes comparable with unity; we consider here only massless ALP, since for a massive ALP the Landau-Pomeranchuk-Migdal (LPM) effect is ever less relevant. Since this phase difference can be estimated in vacuum as $\Phi= (E_a-\bk\cdot\bn)\ell$ over a distance $\ell$, where $\bk$ is the momentum of the ALP and $\bn$ is the direction of the electron, we have for a typical emission angle $\theta\sim m_{\rm th}/E_\bp$ an accumulated phase $\Phi\sim E_a m_{\rm th}^2\ell/E_\bp^2$. Thus the formation length in vacuum is $\ell^{\rm vac}_{\rm form}\sim \gamma^2/E_a$, where $\gamma$ is the Lorentz factor of the electron. However, in a medium, after radiating the electron keeps being deflected by Coulomb scattering off the ions. Such deflections change the direction of the electron only by very small angles; over a length $\ell$ the mean squared angular deflection $\overline{\theta^2}$ follows a typical random-walk fashion with $\overline{\theta^2}\sim \ell/\ell_{\rm Coul}$, where $\ell_{\rm Coul}$ is the mean free path for Coulomb scattering.\footnote{This is computed using the transport cross section $\sigma_{\rm Coul}\sim \alpha^2/E_\bp^2$, so that $\ell_{\rm Coul}^{-1}\sim \sigma_{\rm Coul} n_{\rm ion}$ and over a distance $\ell_{\rm Coul}$ the electron is deflected by angles of order unity.} Thus, this deflection induces an additional dephasing between the electron and the radiated field of the order of $\Phi\sim E_a \overline{\theta^2}\ell\sim E_a \ell^2/\ell_{\rm Coul}$. Thus, in a medium the formation length may be reduced to $\ell_{\rm form}\sim \sqrt{\ell_{\rm Coul}/E_a}$ if the Coulomb mean free path is sufficiently short. In turn, the bremsstrahlung emissivity is reduced by an amount $\sim \ell_{\rm form}/\ell_{\rm form}^{\rm vac}$.

For our case of bremsstrahlung emission from SN, we may initially conclude that the LPM regime is indeed the relevant one. Since the ALP energy is $E_a\sim T$, while $\gamma\sim 1/\sqrt{\alpha}$ and $\ell_{\rm Coul}\sim (\alpha^2 n_{\rm ion}/\mu_e^2)^{-1}\sim (\alpha^2 \mu_e)^{-1}$, we find that $\ell^{\rm vac}_{\rm form}\sim 1/\alpha T$ while $\ell_{\rm form}\sim 1/\alpha \sqrt{T \mu_e}$. Thus, it would appear that in the medium the formation length is indeed reduced by a factor $\sqrt{T/\mu_e}$, which would correspondingly reduce the emissivity.

However, it turns out that the LPM effect is likely less relevant than the above argument suggests. We have found that in the medium the bremsstrahlung happens in a regime of strong screening, in which the typical angles between the radiation and the electron is larger than in vacuum, typically of order $\theta_{\bp\bk}\sim k_D/E_\bp\sim k_D/\mu_e$. Therefore, the correct formation length to compare with, in the absence of LPM scattering, is rather of the order of $\ell^{\rm vac}_{\rm form}\sim E_\bp^2/E_a k_D^2$; since $k_D\sim \sqrt{\alpha \mu_e^3/T}$, we find that $\ell^{\rm vac}_{\rm form}$ and $\ell_{\rm form}$ are actually comparable. This means that the conditions for the LPM effect are just satisfied, presumably implying that the bremsstrahlung emissivity might be affected by a reduction of factors of order unity. A detailed treatment of the LPM effect is well known to be a very complicated task even in the quasi-classical regime, well beyond the scope of this paper. Here we limit ourselves to point out that it can induce a suppression in the bremsstrahlung emissivity, but not by a large factor.

A similar argument might also be applied for the competing Compton emission, for which the radiated ALP appears, as we have seen, at an angle $\theta_{\bp\bk}\sim m_{\rm th}/E_\bp\sim 1/\sqrt{\alpha}$. However, in this case the typical ALP energies are of the order of $E_a\sim \mu_e$. Therefore, one can easily check that the formation length in vacuum and in medium are, once again, comparable. For pair production, where the typical radiated energies are of order $\sim T$, the conditions for LPM are more thoroughly satisfied, so that a reduction by a factor $\sim\sqrt{T/\mu_e}$ can be expected, but pair production is anyway subdominant compared to Coulomb emission.

Finally, we should also distinguish very clearly the LPM effect, which can potentially affect slightly the ALP-electron emission, from the multiple-scattering suppression of ALP emission from nucleons (as discussed, e.g., in Ref.~\cite{Raffelt:1991pw}). In the latter case, the suppression is caused by the nucleon scattering many times and losing coherence in its spin over a timescale comparable with the ALP inverse frequency, which causes an effective damping of the nucleon state. In our case, the spin of the electron over the length scale $\ell_{\rm form}$ is barely changed at all; it is only the relative phase between the electron and the radiated ALP that is affected. Indeed, the LPM effect is not a ``damping'' effect, and cannot be modeled by endowing the intermediate electron with an imaginary energy mimicking a loss of coherence (as is instead the case for the multiple-scattering suppression in ALP emission from nucleons); in this case, the electron remains coherent throughout the emission.

Finally, another source of uncertainty, which also has never been mentioned in previous treatment of this problem, is the presence of electron-electron correlations induced by Coulomb interactions. The latter could induce correlations among the electron occupation numbers in the initial and final state, both in Compton and in bremsstrahlung emission. Such correlations could potentially enhance the emission, by reducing the probability of having the initial and the final state electron close together; in other words, the two-particle distribution would not factorize as the product of the initial and final electron distribution, as we are instead assuming here. The rationale for this choice is that we usually are in a regime in which ionic screening is quite larger than electronic screening, since $T\ll \mu_e$. Therefore, correlations among electrons are maintained over a relatively short distance $\lambda_D\sim k_D^{-1}$, where $k_D\sim \sqrt{4\pi\alpha\mu_e^3/T}$ is the Debye momentum induced by the ions; later we will give the precise value for this quantity, but for this section we only need its order of magnitude. Within a sphere of radius $\lambda_D$ and volume $V\sim \lambda_D^3$, the electrostatic interaction energy of the electrons is of the order of $U_{\rm pot}\sim \alpha n_e^2 V^2/\lambda_D\sim \alpha n_e^2 \lambda_D^5$ while their kinetic energy is $K\sim n_e \mu_e V\sim n_e \mu_e \lambda_D^3$. The ratio between the two is $U/K\sim T/\mu_e$, and therefore in relatively degenerate environments as a supernova core these corrections amount probably to $10\%$. Their general effect tends to enhance the emissivity: in the Compton and bremsstrahlung production, they can reduce the probability of having the initial and final electron close together, while in pair annihilation and pair coalescence they can enhance the probability of having the electrons and positrons close together. However, a discussion beyond these order-of-magnitude arguments must necessarily involve a complex description. In non-relativistic contexts, such as in ALP and photon production via Compton processes in the Sun, Coulomb correlations must be included, since electrons are among the dominant source of screening and the Coulomb corrections are therefore of order unity, and they 
are often included by accounting for the static structure function of the electrons. However, this simple treatment can only be done because the electrons are strongly non-relativistic in the Sun and therefore essentially recoilless and static throughout the Compton process. This is not the case here, where the electron is dynamic during the radiation process. The uncertainties they induce are anyway smaller than the ones connected with the LPM suppression, which are even more complex to include. 

\begin{figure*}
	\centering
	\includegraphics[width=1\textwidth]{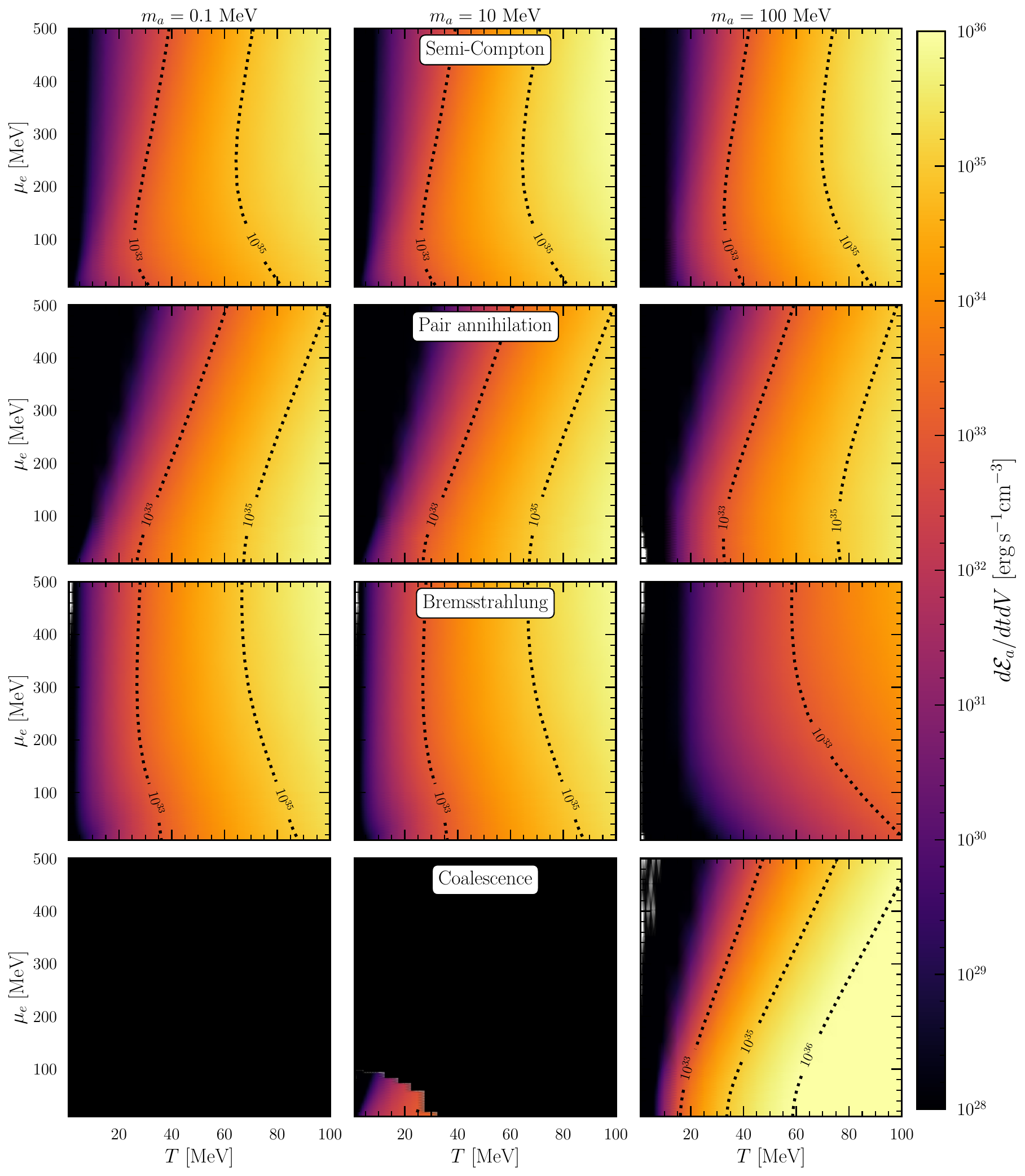}
	\caption{Emissivities for varying temperature and chemical potential, for the four different emission processes. We show the results for three different benchmark values of the ALP mass. The black regions in the coalescence process indicate parameter space where the emissivity vanishes, as $m_a < 2\,m_{\rm th}$.}
	\label{Fig:All_emissivities_contour}
\end{figure*}

\subsection{Electron-positron coalescence}\label{sec:coalescence}

At large ALP masses, the main production channel is direct electron-positron coalescence, $e^++e^-\rightarrow a$, which becomes kinematically allowed. Unlike the other processes, this one is relatively straightforward. The emissivity is given by
\begin{align}
    \frac{dn_a}{dtdV}=&\,\int \frac{d^3\bp}{(2\pi)^3 2E_\bp}\frac{d^3\bl}{(2\pi)^3 2E_\bl}\frac{d^3\bk}{(2\pi)^3 2E_\bk}
    \nonumber\\
    &\times f_{e^-}(E_\bp) f_{e^+}(E_\bl) (2\pi)^4 \delta(p+l-k) |\mathcal{M}|^2.
\end{align}
The squared matrix element, summed over the spins of the electron and positron, and taken for massless leptons, is
\begin{equation}
    |\mathcal{M}|^2=4g_{ae}^2 p\cdot l=2g_{ae}^2 m_a^2.
\end{equation}

After replacing and performing the relevant integrations we find
\begin{equation}
    \frac{dn_a}{dE_\bk dtdV}=\frac{g_{ae}^2 m_a^2}{16\pi^3}\int_{\frac{m_a^2}{2(E_\bk+k)}}^{\frac{m_a^2}{2(E_\bk-k)}}\frac{dp}{(e^{\frac{p-\mu_e}{T}}+1)(e^{\frac{E_\bk-p+\mu_e}{T}}+1)},
\end{equation}
essentially the same result as for neutrino-neutrino coalescence of heavy Majorons~\cite{Fiorillo:2022cdq}. The integral can be done explicitly to give
\begin{align}\label{eq:coalescence_spectrum}
    \frac{dn_a}{dE_a dt dV}&=\frac{g_{ae}^2 m_a^2 T}{16\pi^3(e^{\frac{E_a}{T}}-1)}
    \nonumber\\
    &\times\log\left[\frac{(e^{\frac{m_a^2}{2(E_a-k)T}}+e^{\frac{\mu_e}{T}})(e^{\frac{m_a^2}{2(E_a+k)T}}+e^{\frac{E_a+\mu_e}{T}})}{(e^{\frac{m_a^2}{2(E_a+k)T}}+e^{\frac{\mu_e}{T}})(e^{\frac{m_a^2}{2(E_a-k)T}}+e^{\frac{E_a+\mu_e}{T}})}\right].
\end{align}
Since we have assumed massless leptons, this approximation holds only for $m_a\gtrsim 2m_{\rm th}$, which is anyway the region where coalescence would dominate over the other processes; therefore, we consider vanishing emission from the regions where $m_a<2m_{\rm th}$.

\begin{figure*}
	\centering
	\includegraphics[width=1\textwidth]{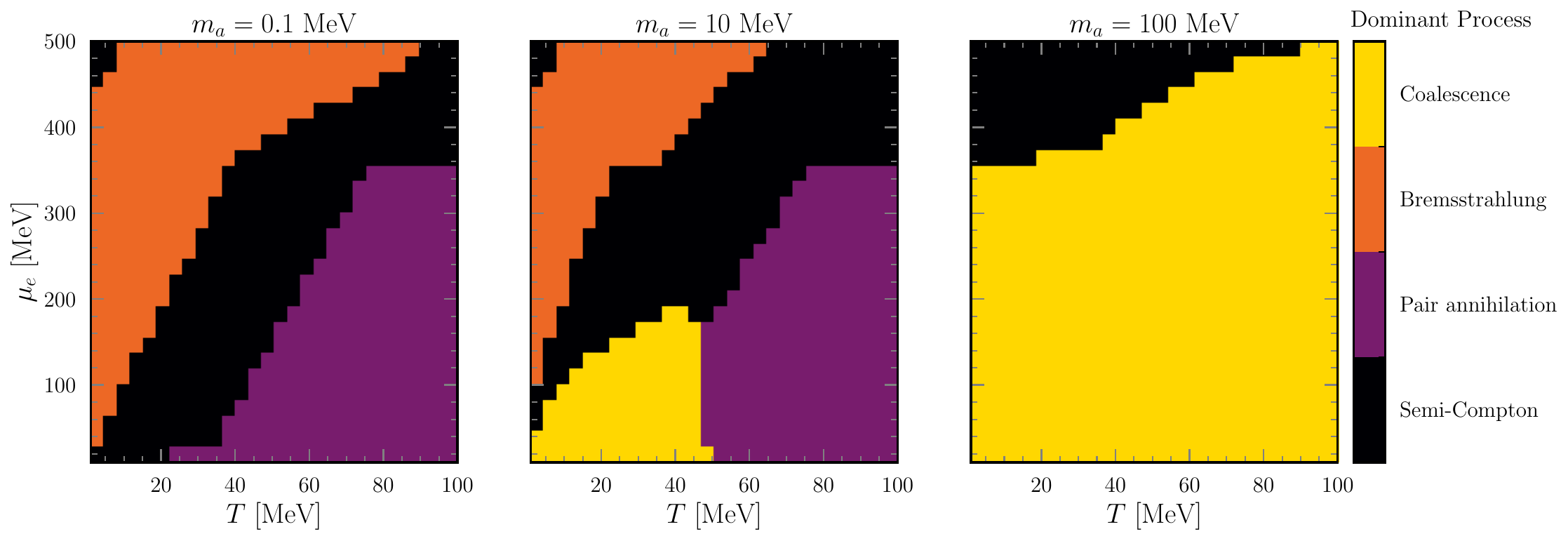}
	\caption{Dominance of the four ALP production channels across the  ($T,\mu_e$)  parameter space for different ALP masses. The dominant process at each point is determined by the highest energy production rate among Semi-Compton scattering, pair annihilation, bremsstrahlung, and electron-positron coalescence.  }
	\label{Fig:Region_plot}
\end{figure*}

\subsection{Processes induced by photon effective mixing}\label{sec:loops}

Even in the absence of a tree-level coupling between the ALP and the photon, the ALP-electron interaction induces a loop-level ALP-photon coupling. It has been argued~\cite{Ferreira:2022xlw} that this induced coupling could open new production channels, potentially dominating over those directly arising from the ALP-electron interaction. These channels include photon-photon coalescence, $\gamma\gamma\to a$, at large ALP masses, and Primakoff production, $\gamma N\to a N$, at small ALP masses. However, we argue that this is not the case.

The key issue is that Ref.~\cite{Ferreira:2022xlw} determines the loop-induced ALP-photon coupling in vacuum, using the electron mass $m_e$. By simple dimensional analysis, the induced ALP-photon coupling, $g_{a\gamma}$, which has units of inverse energy, must scale as $g_{a\gamma} \sim \alpha g_{ae}/m_e$, where the factor $\alpha$ arises from the electron-photon vertices. Since $m_e$ is one to two orders of magnitude smaller than the relevant energy scales in a SN, this leads to a significant enhancement of the emissivity from processes involving this coupling. Physically, this enhancement arises because virtual electrons are easily excited due to their low energy cost, $\sim m_e$, which is much smaller than the typical particle energies in the medium.

However, this also immediately reveals the flaw in the above reasoning. In a medium, the energy cost for exciting a virtual electron is not $m_e$ but rather the thermal mass $m_{\rm th} \gg m_e$. Ref.~\cite{Ferreira:2022xlw} asserts that the vacuum mass should always be used in the loop calculation, citing the argument that, in vacuum perturbation theory, the bare mass enters loop diagrams even when the physical mass is renormalized and enters tree diagrams. However, this claim is incorrect: in vacuum perturbation theory, the bare mass is not an observable quantity and thus cannot enter any physical prediction.

In fact, since tree-level processes require the use of medium-modified dispersion relations---an aspect acknowledged by Ref.~\cite{Ferreira:2022xlw}---loop-level processes should consistently incorporate the same modifications. This can also be argued from fundamental principles: unitarity links the imaginary part of loop diagrams to tree-level processes, while causality relates the imaginary and real parts of these diagrams. If tree-level diagrams require renormalization, so do loop-level diagrams.

More generally, an effective loop calculation must always incorporate a properly resummed propagator. As shown in Sec.~\ref{sec:electrons}, this propagator completely loses memory of the vacuum electron mass, which is negligible compared to other relevant energy scales. Physically, exciting a virtual electron in a medium requires a minimum energy of order $m_{\rm th} \gg m_e$, and neglecting this leads to spurious enhancements in the emissivities. Since all loop-induced emissivities scale as $g_{a\gamma}^2$, they receive an additional suppression factor of $(m_e/m_{\rm th})^2 \sim m_e^2 / (\alpha \mu_e^2) \sim 0.01$ compared to the estimates in Ref.~\cite{Ferreira:2022xlw}. For this reason, we do not include these processes in our calculation of the ALP emissivity.

\begin{figure*}
	\centering
	\includegraphics[width=0.8\textwidth]{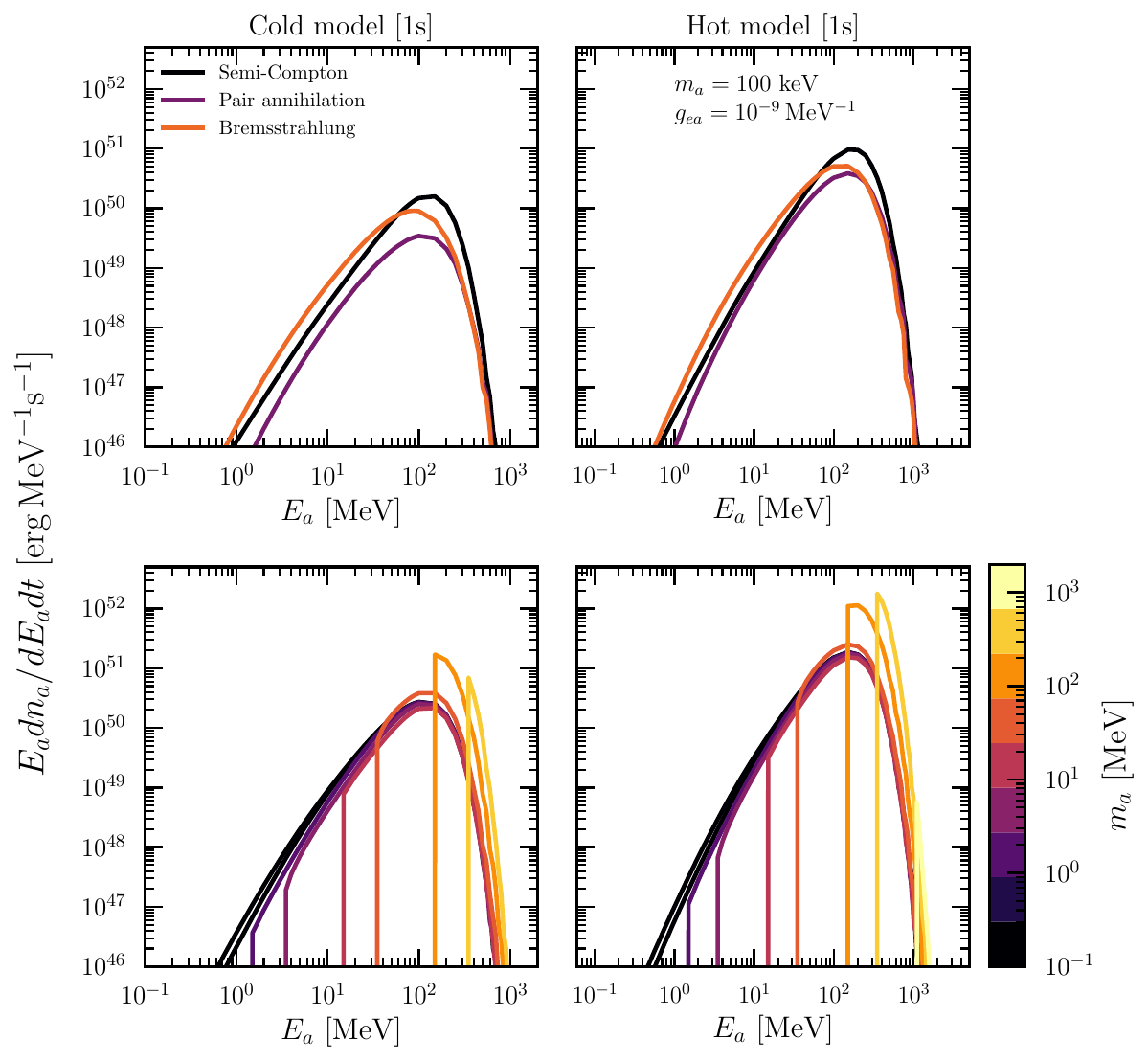}
	\caption{\textit{Upper panel:} Volume-integrated ALP emissivity for $m_a = 100$~keV computed at 1s into the simulation from different production channels for both cold and hot SN models. ALP production is enhanced in the hot model due to its higher temperature profile. The coalescence contribution is zero for such small ALP masses and is therefore not shown. \\
    \textit{Lower panel:} Volume-integrated ALP emissivity as a function of ALP mass, for $g_{ea}=10^{-9}\,\mathrm{MeV^{-1}}$. For small ALP masses, the semi-Compton channel dominates the total emitted energy, while at larger $m_a$, electron-positron coalescence becomes the primary production mechanism.}
	\label{Fig: 2SN models ma=100keV merged}
\end{figure*}

\begin{figure*}
	\centering
	\includegraphics[width=0.9\textwidth]{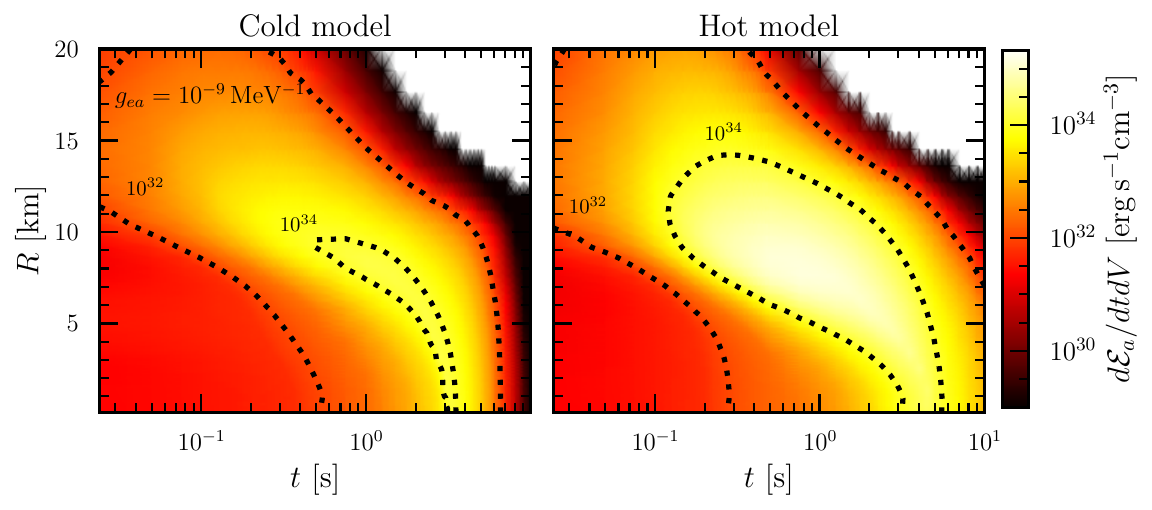}
	\caption{Contours of the volume  and energy integrated emissivity as a function of time and PNS radius for a coupling of $g_{ae} = 10^{-9},\mathrm{MeV}^{-1}$. The emissivity distribution closely traces the temperature evolution within the PNS. The white region corresponds to a chemical potential below 10~MeV, which lies outside the grid range considered in this work.}
	\label{Fig:time vs radius energy integrated emissivity}
\end{figure*}

\section{Mean free path for absorption}
\label{sec:MFP}

For each process by which ALPs can be produced, a corresponding inverse process can lead to their absorption. The mean free path for such absorption is relevant to determine the so-called trapping of the species at large values of the ALP-electron coupling. The principle of detailed balance allows us to obtain the absorption rate directly from the expressions for their spontaneous emission rate. We have previously derived the emission rate for each process as $dn_a/dE_a dt dV$; the corresponding rate of change of the ALP distribution function, due to spontaneous emission alone, is
\begin{equation}
    Q_a=\left(\frac{\partial f_a}{\partial t}\right)_{\rm em}=\frac{2\pi^2}{E_a \sqrt{E_a^2-m_a^2}}\frac{dn_a}{dE_a dt dV}.
\end{equation}

When accounting for stimulated emission, as well as absorption, the kinetic equation for the ALP in a medium in thermal and chemical equilibrium reads
\begin{equation}
    \frac{\partial f_a}{\partial t}=Q_a(1+f_a)-\gamma_a f_a=Q_a-\tilde{\gamma}_a f_a,
\end{equation}
where $\gamma_a$ is the absorption rate of the ALP (we do not use the often-adopted notation $\Gamma_a$, which is usually reserved for the rest-frame, not lab-frame, decay rate of the ALP). The reduced absorption rate $\tilde{\gamma}_a=\gamma_a-Q_a$ is more useful, as it is the physical rate with which the distribution function evolves towards equilibrium. From the principle of detailed balance, we therefore immediately conclude
\begin{equation}\label{eq:detailed_balance}
    \tilde{\gamma}_a=Q_a(e^{E_a/T}-1)=\frac{2\pi^2 (e^{E_a/T}-1)}{E_a\sqrt{E_a^2-m_a^2}}\frac{dn_a}{dE_a dt dV},
\end{equation}
which allows us to determine the absorption rate from the numerical results provided in the accompanying GitHub repository~\cite{AxionRepo}. As a simple consistency check, for the case of pair coalescence we may use Eq.~\eqref{eq:coalescence_spectrum} to obtain in the limit $T\to 0$, $\mu_e\to 0$ the decay rate $\tilde{\gamma}_a=g_{ae}^2 m_a^3/8\pi E_a=\Gamma_a m_a/E_a$, where $\Gamma_a=g_{ae}^2 m_a^2/8\pi$ is the decay rate of an ALP in vacuum.

Eq.~\eqref{eq:detailed_balance}, while exact, is not always suitable for numerical calculations, as it requires the product of a number that can become very small---the emissivity---with a number that can become very large---the reciprocal of the equilibrium distribution function of the ALPs. Therefore, we rather use the exact expressions for each process, which, while requiring us to perform a separate integral for the absorption rate, do not suffer from the same numerical challenges. Therefore, for semi-Compton, we use the expression

\begin{widetext}

\begin{align}\label{eq:compton_absorption}
    \tilde{\gamma}_a&=\frac{2\pi^2}{E_a \sqrt{E_a^2-m_a^2}}\int_0^{+\infty}pdp\int_0^{+\infty}qdq \int_{-1}^{+1}dX \nonumber
    \\
    &\times\frac{-f_e(p) f_\gamma(q) \left[1-f_e(p+q-E_a)\right]+(1-f_e(p))(1+f_\gamma(q))f_e(p+q-E_a)}{512 \pi^6 |\bp+\bq|}4\pi \alpha g_{ae}^2I,
\end{align}
with $I$ from Eq.~\eqref{eq:compton_I_integral}; for pair annihilation, we use

\begin{align}\label{eq:pair_absorption}
    \tilde{\gamma}_a=&\frac{2\pi^2}{E_a \sqrt{E_a^2-m_a^2}}\int_0^{+\infty}pdp\int_0^{+\infty}qdq \int_{-1}^{+1}dX \nonumber
    \\
   &\times \frac{-f_e(p) f_{e^+}(q+E_a-p) \left[1+f_\gamma(q)\right]+f_\gamma(q)(1-f_e(p))(1-f_{e^+}(q+E_a-p))}{512 \pi^6 |\bp-\bq|}4\pi \alpha g_{ae}^2I,
\end{align}
with $I$ from Eq.~\eqref{eq:pair_I_integral}; for bremsstrahlung, we use
\begin{align}
    \tilde{\gamma}_a=\frac{2\pi^2}{E_a \sqrt{E_a^2-m_a^2}}\int \frac{pkl Z^2 \alpha^2 g^2 n_{\rm ion}}{4\pi^4}(f_{e^-}(E_\bl)-f_{e^-}(E_\bp)) dE_\bp
    \int dx[\mathcal{I}_1+\mathcal{I}_2+2\mathcal{I}_3];
\end{align}
and for pair coalescence we use
\begin{equation}
    \tilde{\gamma}_a=\frac{2\pi^2}{E_a\sqrt{E_a^2-m_a^2}}\frac{g_{ae}^2m_a^2 T}{16 \pi^3}\log\left[\frac{(e^{\frac{m_a^2}{2(E_a-k)T}}+e^{\frac{\mu_e}{T}})(e^{\frac{m_a^2}{2(E_a+k)T}}+e^{\frac{E_a+\mu_e}{T}})}{(e^{\frac{m_a^2}{2(E_a+k)T}}+e^{\frac{\mu_e}{T}})(e^{\frac{m_a^2}{2(E_a-k)T}}+e^{\frac{E_a+\mu_e}{T}})}\right].
\end{equation}

In the limit $T\to 0$, this form may be inconvenient, due to the large exponentials in the logarithms; therefore, if $T\ll\mu_e$, we can use
\begin{equation}
    \tilde{\gamma}_a=\frac{2\pi^2}{E_a \sqrt{E_a^2-m_a^2}}\frac{g_{ae}^2 m_a^2}{16\pi^3}\mathrm{min}\left[\sqrt{E_a^2-m_a^2},\frac{m_a^2}{2(E_a-\sqrt{E_a^2-m_a^2})}-\mu_e\right],
\end{equation}
whereas it vanishes for $\mu_e>m_a^2/2(E_a-\sqrt{E_a^2-m_a^2})$ (for an ALP at rest, we immediately see that indeed for $\mu_e>m_a/2$ the decay rate must vanish because the produced electron cannot lie within the Fermi sea).

One can easily check that these expressions naturally satisfy the general detailed balance expression we derived before, yet they do not suffer from numerical instabilities as the original expression.

\end{widetext}
 
\section{Results}
\label{sec:results}

While we have derived approximate analytical expressions for the ALP emissivity in various regimes and for all relevant processes, these approximations are generally valid only over limited energy ranges and typically for massless ALPs. A complete numerical determination of the emissivity is therefore necessary. To this end, we have performed a comprehensive scan of the relevant conditions in dense and hot environments, such as those found in supernovae SNe and NSMs. The resulting production rates are crucial for a consistent determination of the ALP flux from these extreme environments. To facilitate further studies, we have compiled these emissivities in tabular form and made them publicly available in a GitHub repository~\cite{AxionRepo}. In this section, we present the production rates for typical conditions.

Fig.~\ref{Fig: All_emissivities_ma=100KeV} shows the volumetric ALP production rate for a typical temperature $T=30\,{\rm MeV}$ and chemical potential $\mu_e=150\,{\rm MeV}$, similar to the hottest regions of a SN at about $1$~s post-bounce (see Fig.~1 of Ref.~\cite{Caputo:2021rux}, and our later discussion). We compare our numerical results for $m_a=100$~keV, much smaller than any other scale within the SN environment, with the massless approximations we derived in Sec.~\ref{sec:emission}. We find excellent agreement with all analytical approximations within their respective regimes of validity, which serves as a validation test for the convergence of the multi-dimensional numerical integration. The four components are shown together in Fig.~\ref{Fig:All_emissivities_together}, illustrating that the semi-Compton process---previously neglected---is \edit{comparable to bremsstrahlung and, in fact, dominates at the highest energies, comparable with the electron chemical potential}.

We examine the dependence of the emissivity on the ALP mass in Fig.~\ref{Fig:All_emissivities_different_masses}. In the case of semi-Compton scattering, increasing the ALP mass primarily results in a kinematic cutoff at $E_a < m_a$.  For pair annihilation, and especially for pair coalescence, in addition to this cutoff, there is a net increase in the overall ALP emissivity, driven by the enhanced matrix element for larger ALP masses. For bremsstrahlung, besides the cutoff at $E_a < m_a$, increasing the ALP mass leads to a reduction in the overall emissivity normalization. This occurs because the typical electron deflection angle during ALP emission is no longer set by $m_{\rm th}/\mu_e$ (for electrons with typical energies $\mu_e$) or $k_D/\mu_e$ in the strong screening regime, but rather by $m_a/T$ (since ALPs have typical energies of order $T$). Therefore, for $m_a\gtrsim k_D T/\mu_e$, the minimum allowed angle for bremsstrahlung increases, leading to a suppression of the emissivity, which scales as $\propto m_a^{-2}$.

The overall emissivity is shown in Fig.~\ref{Fig:All_emissivities_contour} as a contour plot for varying temperature and chemical potential. For both semi-Compton and bremsstrahlung processes, we find that the emissivity—particularly for small ALP masses—primarily depends on the temperature and is only mildly sensitive to the chemical potential.  This result is, of course, confirmed by our analytical estimates. In contrast, pair annihilation and especially pair coalescence exhibit a much stronger dependence on the chemical potential, which is easy to understand since at large chemical potentials the concentration of positrons that can participate to the reaction is exponentially suppressed.

Finally, Fig.~\ref{Fig:Region_plot} shows the regions of parameter space in which different reactions are dominating the emissivity, for varying values of the ALP mass. Generally, no process contributes over $90\%$ of the emissivity for most of the parameter space, so that all processes should be kept for an accurate determination of the emission. Yet, qualitatively, we find that in the most interesting parameter space for SNe and NSMs, the dominant process is always semi-Compton emission, which has been neglected altogether in previous works. Bremsstrahlung turns out to never dominate the emission; as discussed in our analytical estimates, the production of photons with energy of order $\sim T$ turns out to scale in the same way for bremsstrahlung and semi-Compton emission, except that bremsstrahlung is further suppressed by nucleon degeneracy. At high temperatures, on the other hand, pair annihilation may become the dominant process. In this sense, the pattern of dominance of different emission processes is qualitatively comparable with the analogous one for neutrino emission (see, e.g., Fig.~C.1 of Ref.~\cite{Raffelt:1996wa}).

\section{New constraints from supernovae}
\label{sec:constraints}

Having revisited ALP production via their coupling to electrons, we now turn to deriving constraints from SN observables. These constraints have been previously examined in Refs.~\cite{Carenza:2021pcm,Ravensburg:2023uin}, but as discussed earlier, the dominant production channel--semi-Compton emission--was overlooked. Furthermore, some relevant observables were not considered. We therefore begin by outlining the key observables that can be used to constrain ALP-electron couplings in supernovae.

\subsection{Cooling argument}\label{sec:cooling}

The historical argument to constrain the emission of new particles from SNe is based on the cooling of the protoneutron star (PNS). The general idea is that, as the PNS cools faster due to non-standard emission, the duration of the neutrino burst, as observed from SN~1987A, should be shortened. This argument was verified by a series of simulations including the cooling due to the emission of axions coupling to nucleons in Refs.~\cite{Mayle:1987as,Turner:1987by,Burrows:1988ah,Mayle:1989yx,Burrows:1990pk}. Subsequently, Raffelt proposed a simple analytical criterion based on these simulations~\cite{Raffelt:1996wa}.
However, as recently discussed in Ref.~\cite{Fiorillo:2023frv}, the main factor that sets the duration of the neutrino burst is the convective heat transport within the PNS, which was not included in these simulations, which calls for a re-assessment of the cooling bounds.

As we will see, for ALPs coupling to electrons cooling bounds are anyway not the most sensitive probe; nevertheless, for reference we do show the cooling argument based on the well-known Raffelt criterion, according to which if the non-standard luminosity exceeds the neutrino luminosity at a time of about $1$~s post-bounce the model can be excluded. However, following Ref.~\cite{Fiorillo:2025yzf}, we consider here only the non-standard energy loss from the PNS within a radius of 20~km, and not the entire luminosity. The reason we make this specification is that, for large couplings, the ALPs remain thermally coupled to the medium up to radii much larger than 20~km. In this case, the surface emission at radii far outside the PNS does not correspond to energy subtracted from the PNS, but rather from these outer radii; only the heat flux of ALPs at 20~km is truly extracted from the PNS itself. Therefore, to determine the ALP luminosity at $1$~s, we use the expression
\begin{equation}\label{eq:cooling_luminosity}
    L_a=L_{\rm PNS}^{(1)}-L_{\rm prog}^{(1)}.
\end{equation}
Here $L_{\rm PNS}^{(1)}$ is the luminosity extracted from the PNS from ALPs that escape from it without decaying
\begin{equation}
    L_{\rm PNS}^{(1)}=\int_0^{R_{\rm NS}} dE_a dR \frac{dL_a}{dR dE_a}\int \frac{dx}{2}e^{-\tau(R,R_{\rm NS})};
\end{equation}
here we assume that the ALP is emitted isotropically from the medium with a differential luminosity per unit radius
\begin{equation}
    \frac{dL_a}{dE_adR}=4\pi R^2 E_a \frac{dn_a}{dE_adt dV}
\end{equation}
at an angle $x=\cos\theta$ from the radial direction; we also introduce

    \begin{equation}
    \tau(R,\RNS)=\begin{cases}
        
    \int_R^{\RNS}\frac{\tilde{\gamma}_a r dr}{v_a \sqrt{r^2-R^2\sin^2\theta}},\;x>0 \\
    \left[\int_{R\sin\theta}^R+\int_{R\sin\theta}^{\RNS}\right]\frac{\tilde{\gamma}_a r dr}{v_a \sqrt{r^2-R^2\sin^2\theta}},\;x<0 
    \end{cases}
\end{equation}
as the optical thickness accumulated by the ALP moving with velocity $v_a$ produced at a radius $R<R_{\rm NS}$ and escaping the PNS at a radius $R_{\rm NS}=20$~km.

The second term in Eq.~\eqref{eq:cooling_luminosity} describes the luminosity that the ALPs from outside the PNS deposit inside it through their decay, or more generally absorption: this is

\begin{eqnarray}
    &&L_{\rm prog}^{(1)}=\int_{R_{\rm NS}}^{R_{\rm prog}}dR dE_a \frac{dL_a}{dR dE_a}\\ \nonumber && \int_{-1}^{-\sqrt{1-\frac{R_{\rm NS}^2}{R^2}}}\frac{dx}{2}e^{-\tau_1(R)}(1-e^{-\tau_2}).
\end{eqnarray}

Here 
\begin{equation}
    \tau_1(R)=\int_{\RNS}^R \frac{\tilde{\gamma}_a rdr}{v_a \sqrt{r^2-R^2\sin^2\theta}}
\end{equation}
is the optical depth accumulated from the production point to the point where the ALP enters the PNS, and
\begin{equation}
    \tau_2=2\int_{R\sin\theta}^{\RNS}\frac{\tilde{\gamma}_ardr}{v_a\sqrt{r^2-R^2\sin^2\theta}}
\end{equation}
is the optical depth accumulated along the part of the trajectory within the PNS. In the limit of very large couplings, the two components $L_a^{\rm PNS}$ and $L_a^{\rm ext}$ become nearly equal, and by a numerical treatment it is very challenging to correctly recover the small difference between the two. In this regime, we use the asymptotic expression from radiative transfer theory, and derived in Ref.~\cite{Fiorillo:2025yzf}
\begin{equation}\label{eq:luminosity_radiative_transfer}
    L_a=\frac{-2\RNS^2 \partial T/\partial R}{3\pi T^2}\int dE_a \frac{E_a^3 p_a v_a e^{E_a/T}}{\tilde{\gamma}_a(e^{E_a/T}-1)^2},
\end{equation}
after verifying that it smoothly merges with the numerical results for intermediate couplings where the numerical integration is still stable. Numerically, we implement this prescription by substituting Eq.~\eqref{eq:cooling_luminosity} with Eq.~\eqref{eq:luminosity_radiative_transfer} when the difference $L_a$ becomes smaller than $5\%$ of $L_{\rm PNS}^{(1)}$.

Eq.~\eqref{eq:cooling_luminosity}, in the limit of small couplings, is dominated by the first term with $\tau(R,\RNS)\to 0$, and therefore describes the standard cooling term. On the other hand, for large couplings, it corresponds to a new prescription for the trapping regime introduced in Ref.~\cite{Fiorillo:2025yzf}. We use it to determine the ALP luminosity $L_a$ at 1~s post-bounce for two SN models, which are chosen to bracket a model with a relatively cold and a relatively hot central PNS. These two models, following Refs.~\cite{Bollig:2020xdr, Caputo:2021rux,Fiorillo:2022cdq,Fiorillo:2024upk}, are the Garching 1D models SFHo-18.8 and LS220-s20.0 that were evolved with the {\sc Prometheus Vertex} code with six-species neutrino transport~\cite{JankaWeb}. They span the extremes of a cold and a hot case, reaching internal $T$ of around 40 vs 60~MeV. We will refer to them simply as the cold and hot model respectively. For the cold (hot) model, the ALP luminosity that can be constrained should exceed the neutrino luminosity, namely $4.4\times 10^{52}$~erg/s ($8.3\times 10^{52}$~erg/s).  In the upper panel of Fig.~\ref{Fig: 2SN models ma=100keV merged}, we show the ALP emissivity for $m_a = 100$keV from different production channels. Semi-Compton emission is clearly the dominant process at small ALP masses, and the overall ALP production is enhanced in the hot model due to its higher temperature profile. This trend is more evident in Fig.~\ref{Fig:time vs radius energy integrated emissivity}, where we show contours of the volume and energy integrated emissivity as a function of time and radius for a coupling of $g_{ae} = 10^{-9},\mathrm{MeV}^{-1}$. The emissivity strength closely follows the temperature evolution within the PNS.

In the lower panel of Fig.~\ref{Fig: 2SN models ma=100keV merged}, we show the volume-integrated total ALP emissivity as a function of the ALP mass, evaluated at 1~s post-bounce for both cold and hot SN models. At small ALP masses, the total emitted energy is dominated by the semi-Compton channel, whereas for large $m_a$, $e^+e^-$ coalescence becomes the dominant process (see also Fig.~\ref{Fig:All_emissivities_different_masses}).

\subsection{Energy deposition in low-energy SNe}

For radiatively decaying particles, such as ALPs coupled to electrons with $m_a \gtrsim 1$~MeV—our regime of interest—a more powerful probe is their radiative decay within the progenitor. Such decays deposit energy that could be observable in the explosion; as proposed in Ref.~\cite{Caputo:2022mah}, observations of low-energy SNe can place strong constraints on this scenario.

The energy deposited by the decaying ALPs outside the PNS, assumed to end at a radius $\RNS$, must include the (negative) contribution of the ALPs produced outside the PNS and decaying within the PNS, which are cooling, rather than heating, the progenitor material. Here we follow Ref.~\cite{Fiorillo:2025yzf} and determine the total rate of deposited energy per unit time at each instant as
\begin{equation}
    L_a=L_{\rm PNS}^{(1)}-L_{\rm prog}^{(1)}-L_{\rm PNS}^{(2)}-L^{(2)}_{\rm prog};
\end{equation}
the first two terms have already been discussed in Sec.~\ref{sec:cooling}, while the last terms are the luminosities that the ALPs decaying outside of the progenitor drain from the system, which are clearly not deposited in the star. These are determined as
\begin{equation}
    L_{\rm PNS}^{(2)}=\int_0^{\RNS}dR dE_a \frac{dL_a}{dE_a dR}\int \frac{dx}{2} e^{-\tau(R,\Rprog)},
\end{equation}
where $\Rprog$ is the progenitor radius; following Ref.~\cite{Caputo:2022mah}, we consider both an aggressive ($R_1=5\times 10^{13}$~cm, with a larger progenitor and more energy deposition) and a conservative case ($R_2=3\times 10^{12}$~cm, with a smaller progenitor and less energy deposition). For $L_{\rm prog}^{(2)}$, we use the same form integrated between $\RNS$ and $\Rprog$. Also in this case, when the coupling becomes sufficiently large---at which point $L_{\rm prog}^{(2)}$ and $L_{\rm PNS}^{(2)}$ have essentially vanished---we use Eq.~\eqref{eq:luminosity_radiative_transfer} to determine the luminosity, since a numerical treatment cannot correctly recover the flux from the near-cancellation of the ALPs entering and escaping the PNS. The integrated energy deposited in the progenitor is $\mathcal{E}_a=\int dt L_a$, and the constraints from energy deposition are obtained by requiring $\mathcal{E}_a>10^{50}$~erg for the aggressive choice, corresponding to the radius $R_1$, and $\mathcal{E}_a>10^{51}$~erg for the conservative choice, corresponding to the radius $R_2$.

\subsection{Positrons from $e^+ e^-$ decay}

At relatively small couplings, most ALPs escape the progenitor. Therefore, the most stringent constraints arise from the secondary particles produced in their decays. In this section, we consider the positron population generated by ALPs decaying into $e^+e^-$ pairs after being produced in CCSNe. The resulting signal arises from the cumulative injection of positrons by all SNe occurring in the Galaxy over time.
For these constraints, it suffices to use the time-integrated emissivity over the entire volume of the PNS, which we express as
\begin{equation}
    \frac{d\mathcal{E}_a}{dE_a}=\int \frac{dn_a}{dE_a dV dt} dt dV E_a.
\end{equation}
The ALPs that decay in electron pairs outside the progenitor, with a rest-frame decay rate
\begin{equation}
    \Gamma_{a\to e^+ e^-}=\frac{g_{ae}^2}{8\pi}\sqrt{m_a^2-4m_e^2},
\end{equation}
inject a large amount of positrons, which lose energy through different processes, thermalize, and subsequently annihilate with ambient interstellar electrons at rest, produce a visible signal of a 511 keV line~\cite{Prantzos:2010wi}, which has been measured with the spectrometer on INTEGRAL (SPI)~\cite{Strong:2005zx,Bouchet:2008rp,Siegert:2015knp,Siegert:2019tus,Siegert:2022jii}.

The possibility of constraining novel particles by requiring the resulting 511 keV signal to not exceed the observed data was first proposed in Ref.~\cite{Dar:1986wb}, incidentally a few months before SN~1987A exploded (though it should be noted that this bound depends on the explosion of all past supernovae). Recently, this argument has been rediscovered and applied to several models, including dark photons and sterile neutrinos~\cite{DeRocco:2019njg,Calore:2021lih,Carenza:2023old}. A fraction of the positrons could annihilate before coming to rest, producing a continuum signal at higher energies. This also leads to comparable constraints on novel particles~\cite{DelaTorreLuque:2024zsr}.

The total number of leptons produced in the decay of ALPs outside of the progenitor can be written as
\begin{equation}
    \mathcal{N}_e=\int 2\frac{dE_a}{E_a}\frac{d\mathcal{E}_a}{dE_a}\exp\left[-\frac{\Gamma_a R_2}{v_a}\right],
\end{equation}
where $R_2=5\times 10^{13}$~cm is the typical value of a Type-II SN progenitor radius; we notice in passing that the SN~1987A progenitor, Sanduleak-69 202, was much smaller. For the couplings of interest, the vast majority of the ALPs decay in close proximity to the SN, at distances negligible compared to the Earh-SN separation. The factor $2$ accounts for the two leptons produced per each decaying ALP. Note that in some studies, the bounds from the 511 keV line are expressed in terms of the number of positrons alone. Reference~\cite{DeRocco:2019njg} finds $\mathcal{N}_{e^+}<10\times 10^{52}$ assuming a SN rate of 2 events per century. Reference~\cite{Calore:2021klc} suggests $\mathcal{N}_{e^+}<1.6\times 10^{52}$ by accounting for the angular information of the signal, while Ref.~\cite{Calore:2021lih} suggests $\mathcal{N}_{e^+}<1.4\times 10^{52}$ discussing the uncertainties due to, e.g., the smearing of the signal and the SN rate. A recent analysis finds a bound of $\mathcal{N}_{e^+}<1.8-3\times 10^{52}$~\cite{DelaTorreLuque:2024zsr}. Following Ref.~\cite{Calore:2021lih}, we assume that the amount of injected leptons must be smaller than $\mathcal{N}_e<2.8\times 10^{52}$.

\subsection{Decay into gamma-rays}

While the dominant ALP decay channel is $a\to e^+e^-$, there are still secondary decay channels which involve gamma-ray production. These are phenomenologically more relevant, because the gamma-rays can be directly observed at Earth. 

\subsubsection{Two-photon decay}

The first channel is connected with the two-photon decay $a\to \gamma\gamma$ via the loop-induced ALP-photon coupling. In the mass range $m_a\gtrsim1$~MeV, which is of interest here, the matrix element for $a\to\gamma\gamma$ has an imaginary part, corresponding to the on-shell decay $a\to e^+e^-$ with subsequent pair annihilation $e^+e^-\to\gamma\gamma$. The effective ALP-photon coupling in this range is
\edit{
\begin{equation}\label{eq:loop_induced_coupling}
    g_{a\gamma}=\frac{\alpha g_{ae}}{\pi m_e }\left[-\frac{m_e^2}{m_a^2}\left(\pi+i\log\left(\frac{m_a+\sqrt{m_a^2-4m_e^2}}{m_a-\sqrt{m_a^2-4m_e^2}}\right)\right)^2\right].
\end{equation}
Here we use the loop-induced coupling for a pseudoscalar, rather than a pseudoderivative, axion–electron interaction~\cite{Bauer:2017ris}. This is consistent with our choice in Sec.~\ref{sec:pseudoscalar} to remove the anomalous contribution to the axion–photon coupling, since this contribution persists even in the ultraviolet, arising primarily from fluctuations at arbitrarily high energies. This is equivalent to considering an axion-like particle coupling through a pseudoderivative interaction, but with a tree-level axion-photon coupling which cancels the anomalous one induced by the triangle anomaly.}\footnote{\edit{In any case, as we find here and in the previous version of the paper---where we have used the loop-induced coupling for a pseudoderivative interaction---the loop-induced two-photon decay leads to constraints that are not competitive with the other observables considered here.}}

Note that the imaginary part here accounts for the contribution from the annihilation of on-shell pairs produced from the same ALP. For sufficiently large couplings, the decaying ALP population outside the SN progenitor becomes high enough that annihilations can also occur between on-shell pairs originating from different ALPs. However, since decay bounds are most relevant at relatively small couplings, we neglect this additional contribution. A proper treatment of this effect would require a non-equilibrium field-theoretic approach in the ALP-electron-photon medium formed outside the SN progenitor.

The rest-frame decay rate due to $a\to \gamma\gamma$ is
\begin{equation}
    \Gamma_{a\to \gamma\gamma}=\frac{|g_{a\gamma}|^2 m_a^3}{64 \pi}.
\end{equation}
For SN~1987A, the gamma-rays produced in the ALP decay, with typical energies $E_\gamma\sim 100$~MeV, should have been observed by the Solar Maximum Mission (SMM). At the time of SN~1987A, this reported observations up to a maximum exposure time $t_{\rm max}=223.2$~s. Therefore, we can set constraints by requiring that the total gamma-ray fluence within this time interval at Earth were lower than the upper bounds set by SMM. The gamma-ray fluence is obtained in Appendix~\ref{app:decay_photons}, integrated over the photon frequency range between 25--100~MeV where SMM placed the upper bounds. Our constraints are therefore obtained when the gamma-ray fluence in this energy band exceeds 1.84~cm$^{-2}$ (using a SN-Earth distance of $D=51.2$~kpc), which is the upper bound set by SMM.

\subsubsection{Photons from $a\to e^+ e^-\gamma$ decay}

A second channel for gamma-ray production is the decay $a \to e^+ e^- \gamma$. Since this is a three-body decay, the photon energy is not directly correlated with the emission angle, requiring integration over this angle to determine the photon fluence at Earth. We derive the energy and angular distribution of photons from a single decay in Appendix~\ref{app:decay_channel} and compute the total gamma-ray fluence observed at Earth in Appendix~\ref{app:decay_photons}. Finally, we constrain the ALP coupling by requiring the fluence to remain below the limits set by SMM, as in the case of the two-photon decay.

\begin{figure*}
    \includegraphics[width=\textwidth]{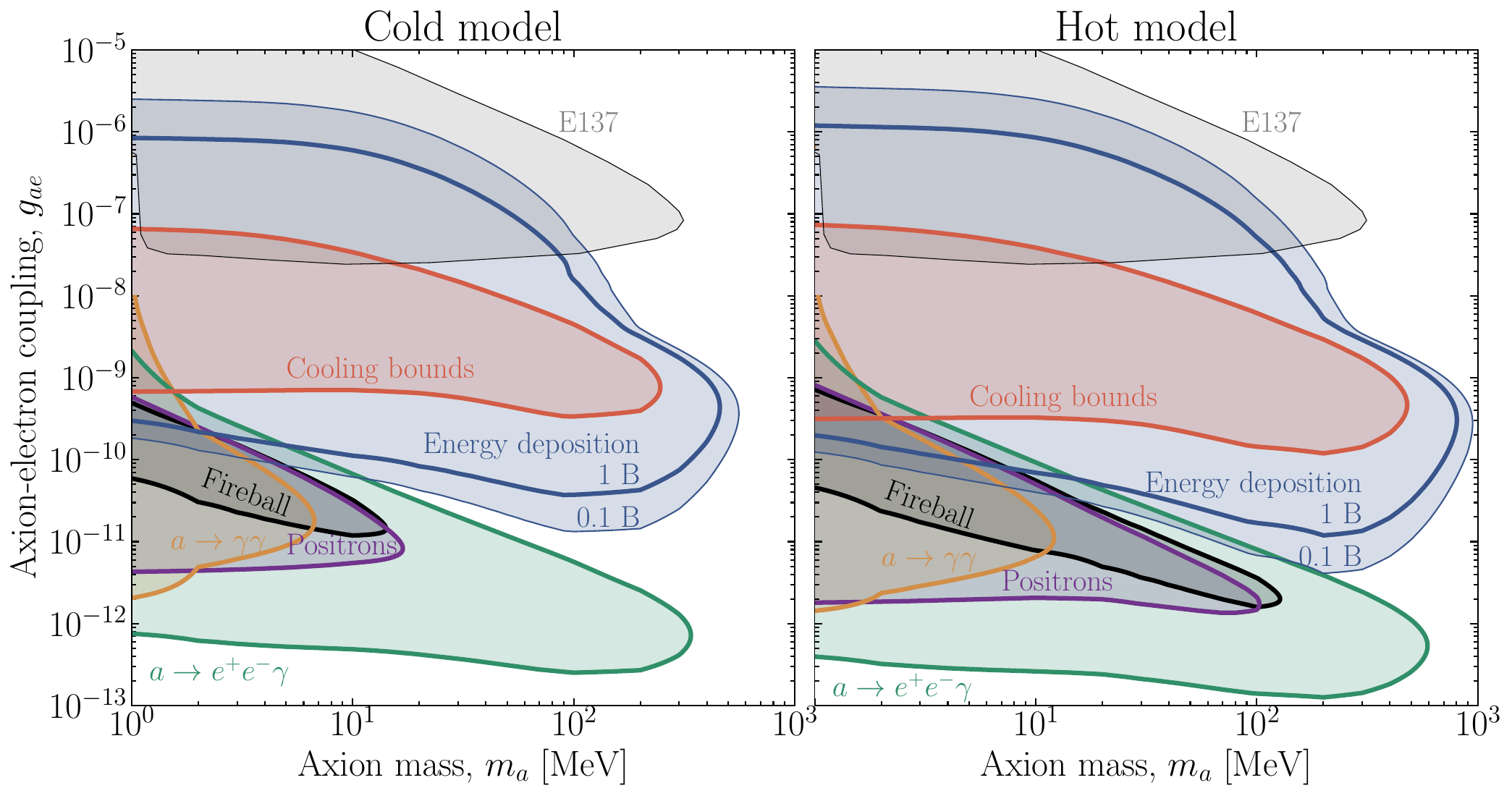}
    \caption{Collected SN constraints on ALP-electron coupling, for the cold model (left) and the hot model (right). We show in gray the beam dump constraints from E137~\cite{Bjorken:1988as} (as shown in Fig.~6 of Ref.~\cite{Liu:2017htz}). We do not show constraints from Orsay~\cite{Davier:1989wz} (as shown in Fig.~4 of Ref.~\cite{Liu:2017htz}), which are obtained for a scalar coupling. In color, we show the new constraints derived in this work.  $\text{B}=\text{Bethe}=10^{51}$~erg.}\label{fig:constraints}
\end{figure*}

\subsection{Fireball formation}

If the ALP-electron coupling is sufficiently large, the pairs and photons produced in ALP decays can reach densities high enough to thermalize in the final state~\cite{Diamond:2023cto}. In the case of ALP-photon coupling, where this effect was first identified, thermalization primarily occurs through pair production followed by bremsstrahlung from the produced pairs. However, in our case, pair production is not required for thermalization, as the ALP decay itself directly generates electron-positron pairs. The only condition for efficient thermalization is that the optical depth for bremsstrahlung within the resulting fireball exceeds unity.

If this condition is met, as shown in Ref.~\cite{Diamond:2023scc}, ALP couplings can still be constrained, but the relevant observable is no longer the gamma-ray flux. Instead, the radiation is reprocessed into X-rays in the sub-MeV region, making the Pioneer Venus Orbiter (PVO) the key experiment for setting constraints. Following Ref.~\cite{Diamond:2023scc}, we identify the parameter space where PVO provides the dominant constraints based on the condition
\begin{equation}
    \frac{2\mathcal{N}}{4\pi r^2}\sigma_{ee\to ee\gamma}>1.
\end{equation}
Here 
\begin{equation}
    \mathcal{N}=\int \frac{dE_a}{E_a}\frac{d\mathcal{E}_a}{dE_a} e^{-\frac{\Gamma_a R_2}{v_a}}
\end{equation}
is the total number of ALPs that decay outside of the progenitor, with the factor $2$ accounting for the two leptons injected from each ALP decay. The radius $r$ at which we evaluate the number density is the average radius at which photons appear outside of the progenitor
\begin{equation}
    r=R_2+\frac{\int \frac{dE_a}{E_a}\frac{d\mathcal{E}_a}{dE_a} e^{-\Gamma_a R_2/v_a}\frac{v_a}{\Gamma_a}}{\mathcal{N}}.
\end{equation}
Finally, the bremsstrahlung cross section is
\begin{equation}
    \sigma_{ee\to ee\gamma}=\frac{8\alpha^3}{m_e^2}\left[\log\left(\frac{m_a}{3m_e e^{\gamma_E}}\right)+\frac{5}{4}\right].
\end{equation}
Here, following the discussion in Ref.~\cite{Diamond:2023scc}, we use Eq.~(16) of Ref.~\cite{Diamond:2021ekg}, where for the rest-frame temperature we use $T=m_a/6$. This is estimated using the fact that the average energy of the leptons produced in the decay, in the rest frame, is $m_a/2$, and after thermalizing these leptons will initially acquire a Maxwell-Boltzmann distribution, since they are significantly underdense compared to their chemical equilibrium with vanishing chemical potential. $\gamma_E$ is the Euler-Mascheroni constant.

\subsection{Comprehensive constraints}

Fig.~\ref{fig:constraints} presents the combined constraints derived from all the arguments discussed above, shown separately for the cold and hot models. The overall structure of the constraints illustrates how different observables probe distinct regimes of decay length.

For relatively short decay lengths and large couplings, the strongest constraints arise from energy deposition inside the star. While cooling bounds have historically been a key consideration, they are generally weaker than energy deposition constraints, as they depend on the non-perturbative impact on the PNS rather than on the exploding material and progenitor. The energy deposition constraints exhibit a characteristic change in slope around 200MeV, marking the transition from ballistic to diffusive energy transport from the PNS to its surroundings, a behavior explored in Ref.~\cite{Fiorillo:2025yzf}.

 For large decay lengths and smaller couplings, the dominant constraints come from the products of ALP decay. Bounds based on the 511~keV line from annihilating positrons and on the two-photon decay $a\to\gamma\gamma$ are generally less competitive, while the most stringent constraints arise from the non-observation of gamma-rays from the decay $a\to e^+ e^- \gamma$. This channel, previously neglected, uniformly provides the strongest constraints. In particular, the lower edge of the bound is very weakly dependent on the mass, something that we expect based on the arguments in Appendix~\ref{app:decay_photons}. Here we also show that the $g_{ae}$ constraints from $a\to\gamma\gamma$ should weaken as $g_{ae} \propto m_a^{-1/2}$, a conclusion that we recover numerically.

The black-shaded fireball region should be understood as the dominant constraint in that parameter space, as fireball formation reprocesses gamma-rays and positrons into X-rays, making all other constraints (except energy deposition) inapplicable in that region, which is instead excluded by the non-observation of X-rays with PVO. We show constraints only for $m_a \gtrsim 1$~MeV, as for lower masses cosmological constraints are by far the dominant ones~\cite{Langhoff:2022bij}. Additional astrophysical bounds from starburst galaxy observations may also be relevant~\cite{Candon:2024eah}. However, most of these constraints--energy deposition, positrons, fireball, and $a\to e^+ e^- \gamma$--depend on decays into electrons and positrons and thus disappear at lower masses. In contrast, constraints from $a\to \gamma\gamma$ persist, scaling as $g_{ae} \propto m_a^{-3/2}$ (see Appendix~\ref{app:decay_photons}), while cooling constraints remain independent of mass even at arbitrarily low $m_a$.

\section{Discussion}
\label{sec:discussion}

We have provided a comprehensive analysis of ALP production processes from electrons under conditions of extreme density and temperature. At this stage, it is useful to summarize our main improvements over previous literature:
\begin{itemize}
    \item We have systematically determined the contribution from semi-Compton processes, which are often considered subdominant due to the degeneracy of the SN environment. Our results show that SN conditions do not suppress semi-Compton production relative to bremsstrahlung, making it the dominant channel in relevant regions.
    \item We consistently account for medium-induced corrections to the electron propagator. The plasmino excitation can be neglected for ultra-relativistic energies, which is always justified since $\alpha \ll 1$. However, this also implies that, to maintain the correct level of precision, medium effects should be omitted in the numerator of the electron propagator rather than treated as an effective mass term. Specifically, the propagator of an electron with effective mass $m_{\rm th}$ is \textit{not} proportional to the usual $\slashed{p} + m_{\rm th}$ factor, a common but incorrect assumption in previous works. While this has little quantitative impact—since the medium energy scale remains small compared to the electrons’ kinetic energy—it is an overlooked conceptual issue.
    \item For bremsstrahlung, we have reduced the four-dimensional numerical integration for the emissivity to a two-dimensional integral, significantly improving computational efficiency and enabling a more complete determination of the emissivity across the relevant parameter space.
    \item We have identified the parameter range in which commonly adopted approximations hold. The most stringent requirement is the assumption of near-massless photons, which allows us to neglect longitudinal plasmon contributions and simplify the kinematics. This assumption requires $T \gg \omega_P$, or equivalently $T \gg \sqrt{\alpha} \mu_e$. In regions of mild degeneracy, where semi-Compton production dominates in SNe, this condition is satisfied, while in strongly degenerate environments where bremsstrahlung dominates, it may fail.
    \item We have identified systematic effects that introduce uncertainties in the emissivity. The most significant is the Landau-Pomeranchuk-Migdal (LPM) effect, which leads to destructive interference in ALP emission due to rapid Coulomb deflections of radiating electrons. This effect is distinct from multiple-scattering suppression relevant for ALP-nucleon couplings. The LPM effect is notoriously difficult to quantify beyond an order-of-magnitude estimate, and while we do not attempt a detailed treatment, we note that it could reduce the emissivity by factors of order unity. No LPM suppression is expected in the large-mass regime, where electron-positron coalescence dominates.
    \item Previous works have suggested that loop-induced ALP-photon couplings might dominate ALP production. We have shown that, when the correct thermal masses are used for electrons in the loop, the resulting emissivity is significantly suppressed and remains subdominant compared to direct ALP-electron interactions.
\end{itemize}

With these improvements, we have systematically determined the ALP emissivity across different production channels over the full relevant range of temperature and electron chemical potential. The resulting emissivity is tabulated and made publicly available in the Github repository~\cite{AxionRepo}, allowing for flexible application to different SN models and exploration of model-dependent systematics.

Finally, we have derived constraints on ALPs with $m_a \gtrsim 1$~MeV from supernovae, a regime not excluded by cosmological constraints. Our main findings are: at low couplings, the strongest constraints come from the decay channel $a \to e^+ e^- \gamma$, which had not been previously considered in this context; at large couplings, energy deposition provides the leading constraint. Other observables---cooling bounds, positron signals, and $a \to \gamma\gamma$---exclude already constrained regions. A portion of the parameter space leads to fireball formation, where gamma-rays and positrons are reprocessed through bremsstrahlung and pair annihilation. In this case, gamma-ray and positron constraints do not apply, but the region is excluded by the upper bounds on X-rays from PVO.

\section*{Note} While our work was in preparation, a preliminary revision of Ref.~\cite{Mueller2020} was presented in a recent talk~\cite{Ravensburg:2024}. The authors of this forthcoming paper include semi-Compton production; however, their estimate of ALP production over a broad mass range is dominated by loop-induced processes, which we argue are significantly suppressed by medium effects. Additionally, the proper treatment of energy deposition constraints requires our newly developed criterion~\cite{Fiorillo:2025yzf}, which substantially alters the resulting constraints. Finally, as in previous literature, they do not include the $e^+ e^- \gamma$ decay channel, which provides the strongest bounds in the small-coupling regime.

\section*{Data Availability}
The data generated in this study, which include tabulated emissivities and absorption rates for semi-Compton, bremsstrahlung, pair-production, and electron-positron coalescence, are available in the GitHub repository~\cite{AxionRepo}.

\acknowledgments

We thank Georg Raffelt for useful conversations. \edit{We also thank Giuseppe Lucente for pointing out our inconsistent use of the loop-induced ALP-photon coupling, which is corrected in this version.} DFGF is supported by the Alexander von Humboldt Foundation (Germany). TP acknowledges support from NSF grant PHY-2020275 (Network
for Neutrinos, Nuclear Astrophysics, and Symmetries (N3AS)).
EV is supported by the Italian MUR Departments of Excellence grant 2023-2027 ``Quantum Frontiers'' and by Istituto Nazionale di Fisica Nucleare (INFN) through the Theoretical Astroparticle Physics (TAsP) project.
Computations for this research were performed on the Pennsylvania State University’s Institute for Computational and Data Sciences’ Roar Collab supercomputer.

\appendix
\begin{widetext}

\section{Angular integrals}\label{sec:angular_integrals}

In this appendix, we introduce a collection of master integrals over the solid angle for the direction $\bn_l$ of integrand functions containing polynomial functions of $\bn_l$ in the numerator and the denominator, which form the basis of the analytical integration for the bremsstrahlung emissivity.
The results can be expressed entirely in terms of two master integrals; denoting by $\ba$ and $\bb$ two generic vectors, and by $a=|\ba|$ and $b=|\bb|$, we have
\begin{equation}
    \phi(\ba)=\int \frac{d\Omega_\bl}{1+\ba\cdot\bn_l}=\frac{2\pi}{a}\log\left[\frac{1+a}{1-a}\right].
\end{equation}
and
\begin{align}
    I(\ba,\bb)&=\int \frac{d\Omega_\bl}{(1+\ba\cdot\bn_l)(1+\bb\cdot\bn_l)} \nonumber
    \\
    &=\frac{4\pi}{\sqrt{(1-\ba\cdot\bb)^2-(1-a^2)(1-b^2)}}\mathrm{arccosh}\bigg[\frac{1-\ba\cdot\bb}{\sqrt{(1-a^2)(1-b^2)}}\bigg].
\end{align}
When the two vectors $\ba$ and $\bb$ are aligned, the function $I(\ba,\bb)$ can be expressed as a simple combination of functions $\phi(\ba)$ using the identity
\begin{equation}
    \int \frac{d\Omega_\bl}{\left(1+\frac{\ba\cdot\bn_l}{\lambda}\right)\left(1+\frac{\ba\cdot\bn_l}{\gamma}\right)}=\frac{\gamma \phi\left(\frac{\ba}{\lambda}\right)-\lambda \phi\left(\frac{\ba}{\gamma}\right)}{\gamma-\lambda}.
\end{equation}
Clearly in order for the integrals not to be singular, we need $a<1$ and $b<1$, so the logarithms are well defined.

From these two integrals, a collection of progressively more complex integrals are easily constructed, which we will need to express the full result of the bremsstrahlung angular integration. From the integral $\phi(\ba)$, we can already define the function
\begin{equation}
    Q(\alpha,\delta;\beta,\gamma)=\int\frac{d\Omega_\bl}{(\alpha-\beta \bn_l\cdot\bn_p+\gamma\bn_l\cdot\bn_k)(\delta-\beta \bn_l\cdot\bn_p+\gamma\bn_l\cdot\bn_k)},
\end{equation}
such that
\begin{equation}
    Q(\alpha,\delta;\beta,\gamma)=\frac{\delta\phi\left(\frac{\gamma\bn_k-\beta\bn_p}{\alpha}\right)-\alpha \phi\left(\frac{\gamma\bn_k-\beta\bn_p}{\delta}\right)}{\alpha\delta(\delta-\alpha)}.
\end{equation}
Another identity that we can use is
\begin{equation}
    \int \frac{d\Omega_\bl \bn_l}{\left(1+\frac{\ba\cdot\bn_l}{\lambda}\right)\left(1+\frac{\ba\cdot\bn_l}{\gamma}\right)}=\frac{\ba \gamma \lambda}{a^2(\gamma-\lambda)}\left[\phi\left(\frac{\ba}{\gamma}\right)-\phi\left(\frac{\ba}{\lambda}\right)\right],
\end{equation}
which allows us to introduce another function
\begin{equation}
    R_{p,k}(\alpha,\delta;\beta,\gamma)=\int \frac{d\Omega_\bl \bn_l\cdot\bn_{p,k}}{(\alpha-\beta \bn_l\cdot\bn_p+\gamma\bn_l\cdot\bn_k)(\delta-\beta \bn_l\cdot\bn_p+\gamma\bn_l\cdot\bn_k)}
\end{equation}
given by
\begin{equation}
    R_{p,k}(\alpha,\delta;\beta,\gamma)=\frac{\bn_{p,k}\cdot(\gamma\bn_k-\beta\bn_p)}{|\gamma\bn_k-\beta\bn_p|^2 (\alpha-\delta)}\left[\phi\left(\frac{\gamma\bn_k-\beta\bn_p}{\alpha}\right)-\phi\left(\frac{\gamma\bn_k-\beta\bn_p}{\delta}\right)\right],
\end{equation}
and similarly $R_k$
These integrals contain denominators which are quadratic polynomials in $\bn_l\cdot\bn_p$ and $\bn_l\cdot\bn_k$; we can now move one step further, and obtain the integrals containing denominators which are cubic polynomials. The key identity we will use is
\begin{equation}
    \int \frac{d\Omega_\bl}{\left(1+\frac{\ba\cdot\bn_l}{\lambda}\right)\left(1+\frac{\ba\cdot\bn_l}{\gamma}\right)\left(1+\bb\cdot \bn_l\right)}=\frac{1}{\gamma-\lambda}\left[\gamma I\left(\frac{\ba}{\lambda},\bb\right)-\lambda I\left(\frac{\ba}{\gamma},\bb\right)\right].
\end{equation}
Through this identity, we can immediately introduce the function
\begin{equation}
    S(\alpha,\delta,\theta;\beta,\gamma,\sigma)=\int \frac{d\Omega_\bl}{(\alpha-\beta \bn_l\cdot \bn_p+\gamma\bn_l\cdot\bn_k)(\delta-\beta \bn_l\cdot \bn_p+\gamma\bn_l\cdot\bn_k)(\theta-\sigma \bn_l\cdot\bn_k)},
\end{equation}
which is given by
\begin{equation}
    S(\alpha,\delta,\theta;\beta,\gamma,\sigma)=\frac{1}{\theta\delta\alpha(\alpha-\delta)}\left[\alpha I\left(\frac{\gamma \bn_k-\beta \bn_p}{\delta},-\frac{\sigma \bn_k}{\theta}\right)-\delta I\left(\frac{\gamma \bn_k-\beta \bn_p}{\alpha},-\frac{\sigma \bn_k}{\theta}\right)\right].
\end{equation}
Through repeated additions and subtractions, we can now obtain
\begin{equation}
    T_k(\alpha,\delta,\theta;\beta,\gamma,\sigma)=\int \frac{d\Omega_\bl \bn_l\cdot\bn_k}{(\alpha-\beta \bn_l\cdot \bn_p+\gamma\bn_l\cdot\bn_k)(\delta-\beta \bn_l\cdot \bn_p+\gamma\bn_l\cdot\bn_k)(\theta-\sigma \bn_l\cdot\bn_k)},
\end{equation}
given by
\begin{equation}
    T_k(\alpha,\delta,\theta;\beta,\gamma,\sigma)=\frac{\theta S(\alpha,\delta,\theta;\beta,\gamma,\sigma)-Q(\alpha,\delta;\beta,\gamma)}{\sigma},
\end{equation}
and 
\begin{equation}
    T_p(\alpha,\delta,\theta;\beta,\gamma,\sigma)=\int \frac{d\Omega_\bl \bn_l\cdot\bn_p}{(\alpha-\beta \bn_l\cdot \bn_p+\gamma\bn_l\cdot\bn_k)(\delta-\beta \bn_l\cdot \bn_p+\gamma\bn_l\cdot\bn_k)(\theta-\sigma \bn_l\cdot\bn_k)},
\end{equation}
given by
\begin{equation}
    T_p(\alpha,\delta,\theta;\beta,\gamma,\sigma)=\frac{S(\alpha,\delta,\theta;\beta,\gamma,\sigma)(\alpha\sigma+\theta\gamma)}{\beta\sigma}-\frac{\gamma Q(\alpha,\delta;\beta,\gamma)}{\beta \sigma}-\frac{1}{\beta\delta\theta}I\left[\frac{\gamma\bn_k-\beta\bn_p}{\delta},-\frac{\sigma \bn_k}{\theta}\right].
\end{equation}
Finally, to determine the bremsstrahlung emissivity, we will also need integrals in which the denominator is a quartic polynomial of a special form; thse integrals have the form
\begin{equation}
    P(\alpha,\delta,\theta;\beta,\gamma,\sigma)=\int \frac{d\Omega_\bl}{(\alpha-\beta \bn_l\cdot \bn_p+\gamma\bn_l\cdot\bn_k)(\delta-\beta \bn_l\cdot \bn_p+\gamma\bn_l\cdot\bn_k)(\theta-\sigma \bn_l\cdot\bn_k)^2},
\end{equation}
and
\begin{equation}
    F_{k,p}(\alpha,\delta,\theta;\beta,\gamma,\sigma)=\int \frac{d\Omega_\bl \bn_l\cdot \bn_{k,p}}{(\alpha-\beta \bn_l\cdot \bn_p+\gamma\bn_l\cdot\bn_k)(\delta-\beta \bn_l\cdot \bn_p+\gamma\bn_l\cdot\bn_k)(\theta-\sigma \bn_l\cdot\bn_k)^2}.
\end{equation}
These integrals are easily obtained by the identity $P(\alpha,\delta,\theta;\beta,\gamma,\sigma)=-\partial S(\alpha,\delta,\theta;\beta,\gamma,\sigma)/\partial\theta$; $F_{k,p}$ are related by analogous identities to $T_{k,p}$.

\section{Kernels for bremsstrahlung emissivity}\label{sec:kernels}

As explained in the main text, the bremsstrahlung emissivity reduces to a single numerical integral of three kernel functions $\mathcal{I}_1$, $\mathcal{I}_2$, and $\mathcal{I}_3$. Here we provide explicit expressions for these kernels in terms of the elementary functions introduced in Appendix~\ref{sec:angular_integrals}.

The simplest integral is $\mathcal{I}_1$
\begin{equation}
    \mathcal{I}_1=\int d\Omega_\bl \frac{\mathcal{T}_1}{D_p^2 q^2 (q^2+k_D^2)}.
\end{equation}
The denominator is composed of the factors $D_p^2$, which does not depend on $\Omega_\bl$, and
\begin{align}
    q^2 (q^2+k_D^2)=\, &(p^2+l^2+k^2-2pk x-2pl \bn_l\cdot\bn_p+2kl \bn_l\cdot\bn_k)
    \nonumber \\
    &\times
    (p^2+l^2+k^2+k_D^2-2pk x-2pl \bn_l\cdot\bn_p+2kl \bn_l\cdot\bn_k),
\end{align}
which clearly has the same form as the denominators of the integrals $Q$ and $R_{p,k}$ introduced in Appendix~\ref{sec:angular_integrals}. The numerator is a linear function of $\bn_l\cdot\bn_p$ and $\bn_l\cdot \bn_k$, so it is easy to recognize a linear combination of these integrals in the form
\begin{equation}
    \mathcal{I}_1=-\frac{\left[l p(m_a^2-2E_\bk^2)+2 E_\bk p k l x\right] Q+m_{a}^2 l p R_p+\left[2l E_\bk p( E_\bk x-k)-2m_a^2 l p x\right]R_k}{(m_a^2-2E_\bp E_\bk+2pkx)^2},
\end{equation}
where the functions $Q$, $R_p$, and $R_k$ are all evaluated at $\alpha=p^2+l^2+k^2-2pkx$, $\delta=p^2+l^2+k^2+k_D^2-2pkx$, $\beta=2pl$, and $\gamma=2kl$.

The next integral, in order of complexity, is $\mathcal{I}_3$
\begin{equation}
    \mathcal{I}_3=\int d\Omega_\bl \frac{\mathcal{T}_3}{D_p D_l q^2 (q^2+k_D^2)};
\end{equation}
in this case, the denominator $D_l$ is a linear polynomial in $\bn_l\cdot \bn_k$, so the integrals are recognized of the same form as the functions $S$ and $T_{p,k}$ from Appendix~\ref{sec:angular_integrals}. By matching the coefficients in the numerator, we can write
\begin{equation}
    \mathcal{I}_3=-\frac{m_a^2 l p S+lp(m_a^2-2E_\bk^2)T_p+2lpk^2 x T_k}{m_a^2-2E_\bp E_\bk+2pkx},
\end{equation}
where the functions $S$, $T_p$, and $T_k$ are all evaluated at $\alpha=p^2+l^2+k^2-2pkx$, $\delta=p^2+l^2+k^2+k_D^2-2pkx$, $\theta=m_a^2+2 E_\bl E_\bk$, $\beta=2pl$, $\gamma=2kl$, and $\sigma=2lk$ (notice that in the denominator we must keep the non-vanishing electron thermal mass, and therefore $\theta$ depends on $E_\bl$, not on $l$).

Finally, for $\mathcal{I}_2$
\begin{equation}
    \mathcal{I}_2=\int d\Omega_\bl \frac{\mathcal{T}_2}{D_l^2 q^2 (q^2+k_D^2)}.
\end{equation}
Since $D_l$ now appears squared, the integral can now be reduced to a combination as $\mathcal{I}_3$ for the integrals $P$ and $F_{k,p}$
\begin{equation}
    \mathcal{I}_2=\left[2plE_\bk(E_\bk+kx)-plm_a^2\right] P-m_a^2 l p F_p-2pkl(E_\bk+kx)F_k.
\end{equation}

\section{Decay channels for heavy ALPs}\label{app:decay_channel}

ALPs with masses larger than $m_a\gtrsim 1$~MeV can decay to electrons. Their signatures, however, depend dramatically on their decay channel. The decay to electrons $a\to e^+ e^-$ is the primary channel; however, decays of the form $a\to e^+ e^- \gamma$ or $a\to \gamma\gamma$, even if subdominant, can be relevant for their gamma-ray signatures, since the pure electron-positron decay may not produce a visible signature in all cases. In this section we explicitly obtain the decay rates for all these processes, so as to compare them.

Assuming an interaction Lagrangian
\begin{equation}
    \mathcal{L}_{\rm int}=-ig_{ae}\bar{e} \gamma^5 e a -\frac{1}{4}g_{a\gamma}F_{\mu\nu}\tilde{F}^{\mu\nu}a,
\end{equation}
the decay rate for $a\to e^+ e^-$ in the ALP rest frame is
\begin{equation}
    \Gamma_{a\to e^+ e^-}=\frac{g_{ae}^2}{8\pi}\sqrt{m_a^2-4m_e^2}
\end{equation}
where $m_e$ is the electron mass. The decay rate for $a\to\gamma\gamma$ is
\begin{equation}
    \Gamma_{a\to\gamma\gamma}=\frac{|g_{a\gamma}|^2 m_a^3}{64\pi}.
\end{equation}

In this work, we assume that the ALP does not couple to the photon at high energy scales, and only consider the loop-induced ALP-photon coupling in Eq.~\eqref{eq:loop_induced_coupling}.

Finally, let us determine the photon spectrum produced in the decay $a\to e^+ e^-\gamma$. We denote the four-momenta of the photon, the electron, and the positron, as $k^\mu$, $p^\mu$, and $q^\mu$ respectively. In the ALP rest frame, let us denote by $\bk'$ and $\bp'$ the spatial momenta of the photon and electron respectively; for the positron, $\bq'=-\bk'-\bp'$. We will consider $m_a\gg m_e$, so that $|\bp'|\sim |\bq'|\gg m_e$, although we will later need to consider the corrections due to the electron masses. The number of decays per unit time in the ALP rest frame is then
\begin{equation}
    \frac{dN}{dt'}=\frac{1}{2m_a}\int \frac{d^3\bp'}{(2\pi)^3 2|\bp'|}\frac{d^3\bq'}{(2\pi)^3 2|\bq'|}\frac{d^3\bk'}{(2\pi)^3 2|\bk'|}(2\pi)^4\delta^{(4)}(p+q+k-P) |\mathcal{M}|^2,
\end{equation}
where $P$ is the four-vector of the ALP at rest. After integrating the delta functions, we finally obtain
\begin{equation}\label{eq:spectrum_integral}
    \frac{dN}{dk'dt'}=\frac{1}{64\pi^3m_a}\int_{p'_-}^{p'_+} dp' |\mathcal{M}|^2,
\end{equation}
where for convenience we denote by $k'=|\bk'|$ and $p'=|\bp'|$.
The photon distribution in the lab frame can be obtained by performing the appropriate Lorentz boost; since the photon direction is isotropically distributed, if we denote by $x=\cos\theta_{\bk,\bP}$ the cosine of the angle between the ALP and the photon direction, we then find
\begin{equation}
    \frac{dN}{dk dx dt}=\frac{m_a^2}{2E_a^2(1-v_a x)}\left.\frac{dN}{dk'dt'}\right|_{k'=\frac{kE_a(1-v_a x)}{m_a}},
\end{equation}
where $E_a$ is the ALP energy in the laboratory frame, and $v_a$ its velocity. 
The integral over $p'$ must be performed over the kinematically allowed range; to the lowest non-vanishing order in $m_e$, this implies
\begin{equation}
    p'_-=\frac{m_a}{2}-k'-\frac{m_e^2(m_a-k')}{m_a(m_a-2k')},\qquad p'_+=\frac{m_a}{2}-\frac{m_e^2(m_a-k')}{m_a(m_a-2k')}.
\end{equation}
Since $k'\sim m_a$, these expressions are valid for $m_a\gg m_e$, which is the regime to which we are going to stick. This procedure allows to express the photon spectrum in a three-body process independently of the specifics of the emission process itself, so it applies e.g. to a dark photon decaying to $e^+ e^- \gamma$.

The matrix element, squared and summed over spins and polarizations of the final-state particles, is
\begin{equation}
    |\mathcal{M}|^2=4\pi\alpha g_{ae}^2\left[\frac{\mathcal{T}_1}{D_p^2}+\frac{\mathcal{T}_2}{D_q^2}+\frac{2\mathcal{T}_3}{D_p D_q}\right],
\end{equation}
where $D_p=2p\cdot k$ and $D_q=2q\cdot k$, so we have explicitly separated the contributions from the two Feynman diagrams and their interference. The trace terms are
\begin{eqnarray}
    \mathcal{T}_1&=&8\left[2(p+k)\cdot q (p+k)\cdot p-(p+k)^2 p\cdot q\right],    \nonumber
    \\
 \mathcal{T}_2&=&8\left[2(q+k)\cdot q (q+k)\cdot p-(q+k)^2 p\cdot q\right],\\
    \nonumber \mathcal{T}_3&=&16(p+k)\cdot q(q+k)\cdot p.
\end{eqnarray}

We finally reach the final expression, in which we must keep the lowest-order corrections due to $m_e$ in the denominator in order to regularize the enhancement for very soft particles emitted
\begin{eqnarray}
    |\mathcal{M}|^2=16\pi\alpha g_{ae}^2 \left[\frac{m_a-2p'}{2p'+2k'-m_a+\frac{m_e^2}{p'}}+\frac{2p'+2k'-m_a}{m_a-2p'-\frac{m_e^2}{p'}}-\frac{8p'(p'+k'-m_a)}{\left(m_a-2p'-\frac{m_e^2}{p'}\right)\left(2p'+2k'-m_a+\frac{m_e^2}{p'}\right)}\right].
\end{eqnarray}

The corrections due to $m_e$ in the denominator are important only when $p'$ is very close to the values that would otherwise make it vanish, i.e. $p'=m_a/2-k'$ in the first denominator and $p'=m_a/2$ in the second denominator; therefore, we may approximately replace this expression by
\begin{align}
    |\mathcal{M}|^2=16\pi\alpha g_{ae}^2 \Bigg[\frac{m_a-2p'}{2p'+2k'-m_a+\frac{2m_e^2}{m_a-2k'}}&+\frac{2p'+2k'-m_a}{m_a-2p'-\frac{2m_e^2}{m_a}}
    \nonumber
    \\  &-\frac{8p'(p'+k'-m_a)}{\left(m_a-2p'-\frac{2m_e^2}{m_a}\right)\left(2p'+2k'-m_a+\frac{2m_e^2}{m_a-2k'}\right)}\Bigg].
\end{align}

The expression can be simplified by isolating the terms that would be logarithmically divergent in the limit $m_e\to 0$, obtaining
\begin{equation}
    |\mathcal{M}|^2=16\pi\alpha g_{ae}^2\left[\frac{2k'}{2p'+2k'-m_a+\frac{2m_e^2}{m_a-2k'}}+\frac{2k'}{m_a-2p'-\frac{2m_e^2}{m_a}}+\frac{2m_a(m_a-2k')}{\left(2p'+2k'-m_a+\frac{2m_e^2}{m_a-2k'}\right)\left(m_a-2p'-\frac{2m_e^2}{m_a}\right)}\right].
\end{equation}

The integral in Eq.~\eqref{eq:spectrum_integral} can now be expressed in terms of two elementary integrals which are in fact identical to lowest order in $m_e^2$
\begin{eqnarray}
    I=\int_{p'_-}^{p'_+}\frac{dp'}{2p'+2k'-m_a+\frac{2m_e^2}{m_a-2k'}}=\int_{p'_-}^{p'_+}\frac{dp'}{m_a-2p'-\frac{2m_e^2}{m_a}}=\frac{1}{2}\log\left[\frac{m_a(m_a-2k')}{m_e^2}\right],
\end{eqnarray}
so that
\begin{equation}
    \frac{dN}{dk'dt'}=\frac{\alpha g_{ae}^2}{4\pi^2m_a}\log\left(\frac{m_a(m_a-2k')}{m_e^2}\right)\left[2k'+\frac{m_a(m_a-2k')}{k'}\right].
\end{equation}
This expression is of course not rigorously valid when $m_a-2k'\sim m_e^2/m_a$, when the energy available for electrons and positrons is much smaller than $m_a$ and therefore the assumption $p'\gg m_e$ breaks down. However, this is only a very narrow portion of the energy interval. Instead, for $k'\to 0$, the production rate diverges as $k'^{-1}$, which corresponds to the well-known soft divergence for emission of massless photons. We finally find the photon production rate in the lab frame as
\begin{align}\label{eq:distribution_eplemiphot}
    \frac{dN}{dE_\gamma dxdt}=\frac{\alpha g_{ae}^2\left[m_a^4-2E_a E_\gamma m_a^2 (1-v_a x)+2E_a^2E_\gamma^2(1-v_a x)^2\right]}{8\pi^2 E_\gamma E_a^3(1-v_ax)^2}
    \\
    \nonumber \times\log\left[\frac{m_a^2-2E_a E_\gamma(1-v_a x)}{m_e^2}\right] \Theta\left[m_a-\frac{2E_\gamma E_a(1-v_a x)}{m_a}\right],
\end{align}
where for massless photons we have replaced $k=E_\gamma$, and we have explicitly introduced the kinematic condition $m_a>2k'$ through a Heaviside theta.

For comparison, the photon production rate due to the loop-induced two-photon decay is
\begin{equation}
    \frac{dN}{dE_\gamma dxdt}=\frac{2\Gamma_{a\to\gamma\gamma}m_a}{E_a^2 v_a}\delta\left(x-\frac{2E_\gamma-m_a^2/E_a}{2E_\gamma v_a}\right) \Theta\left(\frac{E_a(1+v_a)}{2}-E_\gamma\right)\Theta\left(E_\gamma-\frac{E_a(1-v_a)}{2}\right),
\end{equation}
where the factor $2$ comes from the two photons emitted, and we have explicitly marked with the Heaviside theta functions the range in which $E_\gamma$ is comprised. Therefore, we get
\begin{equation}\label{eq:distribution_twophotons}
    \frac{dN}{dE_\gamma dxdt}=\frac{g_{a\gamma}^2 m_a^4}{32\pi E_a^2 v_a}\delta\left(x-\frac{2E_\gamma-m_a^2/E_a}{2E_\gamma v_a}\right) \Theta\left(\frac{E_a(1+v_a)}{2}-E_\gamma\right)\Theta\left(E_\gamma-\frac{E_a(1-v_a)}{2}\right).
\end{equation}
Since this is a two-body decay, rather than a three-body one, there is a one-to-one relation between the energy of the photon and its emission angle.

\section{Decay photons from ALPs}\label{app:decay_photons}

The photons observed at Earth have a specific time and angular structure, which we now determine. Our main simplifying assumptions will be: \textit{i)} ALPs decay sufficiently far away from the protoneutron star (PNS) that they can be considered radial; \textit{ii)} ALPs decay over a length scale much shorter than the distance between the supernova (SN) and the Earth, which implies that the photons detected at Earth come essentially from the direction of the SN; \textit{iii)} the duration of the photon signal observed at Earth is much longer than the time over which the ALPs are emitted, so that their emission can be considered instantaneous.

Because of the assumption \textit{i)}, we can write the kinetic equation for the photon distribution $f_\gamma=dN_\gamma/dVdx$, where $x=\cos\theta$ is the cosine of the angle between the photon and the radial direction, as
\begin{equation}
    \frac{\partial f_\gamma}{\partial t}+x\frac{\partial f_\gamma}{\partial r}+\frac{1-x^2}{r}\frac{\partial f_\gamma}{\partial x}=\int dE_a \frac{dN_a}{dE_a} \exp\left[-\frac{\Gamma_a m_a r}{p_a}\right]\frac{1}{4\pi r^2 v_a}\frac{dN_\gamma}{dE_\gamma dx dt} \delta\left(t-\frac{r}{v_a}+D\right),
\end{equation}
where $dN_\gamma/dE_\gamma dx dt$ is the photon distribution produced in the decay obtained in Sec.~\ref{app:decay_channel}, $dN_a/dE_a$ is the differential number of ALPs emitted by the SN over the entire duration of the event, and $\Gamma_a$ is the total decay rate (in the ALP rest frame) for an ALP with energy $E_a$. Here, because of assumption \textit{iii)}, all the ALPs are produced at a fixed instant $t_{\rm em}=-D$, where $D$ is the Earth-SN distance (so we measure times with $t=0$ coinciding with the detection of the neutrino burst at Earth). This equation can be integrated to provide the photon distribution function at Earth
\begin{align}
    f_\gamma=\int dE_a\int_{-1}^x \frac{dx'}{4\pi r v_a \sqrt{1-x^2}\sqrt{1-x^{'2}}} &\frac{dN_a}{dE_a} \frac{dN_\gamma}{dE_\gamma dx' dt}
    \nonumber
    \\
   &\times \exp\Bigg({-\frac{\Gamma_a m_a r \sqrt{1-x^{2}}}{p_a \sqrt{1-x^{'2}}}}\Bigg)\delta\left[t-r x-\frac{r\sqrt{1-x^2}}{v_a\sqrt{1-x^{'2}}}\left(1-v_a x'\right)+D\right].
\end{align}
Here $x'$ is the cosine of the angle between the photon and the radial direction at the position of the decay, so it coincides with the angle between the ALP and photon direction at decay.
Because of assumption \textit{ii)}, the distribution function at Earth will be strongly peaked around $x\simeq 1$; we may therefore integrate this equation over $x$ to obtain the photon flux $d\Phi_\gamma/dE_\gamma=\int f_\gamma dx$. We need to evaluate the resulting expression at the Earth, i.e. at $r=D$. The integration over the delta function imposes the condition
\begin{equation}\label{eq:decay_radius}
    r_{\rm dec}=\frac{r\sqrt{1-x^2}}{\sqrt{1-x^{'2}}}=\frac{v_a t}{1-v_a x'};
\end{equation}
notice that $r_{\rm dec}$ coincides with the radius at which the ALP decay.
\begin{equation}
    \frac{d\Phi_\gamma}{dE_\gamma}=\int dE_a\int \frac{dx'}{4\pi r^2(1-v_a x')}\frac{dN_a}{dE_a}\frac{dN_\gamma}{dE_\gamma dx' dt}\exp\left[-\frac{\Gamma_a m_a t}{E_a(1-v_a x')}\right];
\end{equation}
when expressed in terms of the angle between the ALP and the photon direction in the rest frame, this expression coincides with the result obtained in Ref.~\cite{Oberauer:1993yr}. To proceed further, we can now integrate over the relevant exposure time of the experiment. The Solar Maximum Mission (SMM) reported observations up to a maximum exposure time $t_{\rm max}=223.2$~s. On the other hand, there is a minimum exposure time, due to the fact that if the ALP decays within the SN progenitor its produced photons are unobservable; therefore, the minimum exposure time is determined from the condition $r_{\rm dec}>R_2=3\times 10^{12}$~cm (this is the same $R_2$ which served as an upper bound on the decay position for the energy deposition bounds). Therefore, the photon fluence $d\mathcal{F}_\gamma/dE_\gamma=\int dt d\Phi_\gamma/dE_\gamma$ that must be compared with the experimental upper bounds is
\begin{equation}
    \frac{d\mathcal{F}_\gamma}{dE_\gamma}=\int dE_a \int dx' \frac{E_a}{4\pi r^2 \Gamma_a m_a}\frac{dN_a}{dE_a}\frac{dN_\gamma}{dE_\gamma dx' dt} \left[\exp\left(-\frac{\Gamma_a m_a R_2}{p_a}\right)-\exp\left(-\frac{\Gamma_a m_a t_{\rm max}}{E_a(1-v_a x')}\right)\right].
\end{equation}
We notice that the result obtained here is different from the one reported in Ref.~\cite{Jaeckel:2017tud}, in which the request that the ALP decays outside of the progenitor mantle was factorized compared to the probability of producing a photon within the time window of exposure of the experiments. Instead, the above reasoning makes it clear that the two requests hang together, since the position at which the ALP decays is directly connected with the time at which the resulting photons are observed from Eq.~\eqref{eq:decay_radius}. 

For the $a\to e^+e^-\gamma$, the photon spectrum must finally be obtained using the complete two-dimensional distribution in Eq.~\eqref{eq:distribution_eplemiphot}. In order to interpret the numerical results for the constraints, it is useful to understand the expected asymptotic behavior. For low couplings, such that the typical delay is comparable with $t_{\rm max}$, we can expand the exponential
\begin{equation}
    \frac{d\mathcal{F}_\gamma}{dE_\gamma}=\int dE_a \int dx' \frac{t_{\rm max}}{4\pi r^2 (1-v_a x')}\frac{dN_a}{dE_a}\frac{dN_\gamma}{dE_\gamma dx' dt}.
\end{equation}
 From Eq.~\eqref{eq:distribution_eplemiphot}, we see that the order of magnitude of the numerator is approximately the same as the last term in the square bracket, since the largest kinematically allowed values of $E_a E_\gamma (1-v_a x) \sim m_a^2$. Therefore, the distribution $dN/dE_\gamma dx dt$ becomes approximately
\begin{equation}
     \frac{dN}{dE_\gamma dxdt}=\frac{\alpha g_{ae}^2E_\gamma}{4\pi^2 E_a}\log\left[\frac{m_a^2-2E_a E_\gamma(1-v_a x)}{m_e^2}\right] \Theta\left[m_a-\frac{2E_\gamma E_a(1-v_a x)}{m_a}\right];
\end{equation}
this is only a very coarse approximation, that we do not use for practical purposes, but it serves to highlight conceptually the scalings of the resulting emission.
From this approximate form we gather that the photon fluence at Earth from $e^+ e^- \gamma$ is essentially independent of the ALP mass, except for logarithmic terms; therefore, the corresponding constraints on $g_{ae}$ are nearly independent of the mass $m_a$.

For the two-photon decay $a\to \gamma\gamma$, we can proceed one step further and integrate the angle $x'$ analytically using the delta function in Eq.~\eqref{eq:distribution_twophotons}, obtaining
\begin{equation}
    \left(\frac{d\mathcal{F}_\gamma}{dE_\gamma}\right)_{a\to\gamma\gamma}=\int dE_a \frac{g_{a\gamma}^2 m_a^3}{128\pi^2 r^2 \Gamma_a p_a}\frac{dN_a}{dE_a}\left[\exp\left(-\frac{\Gamma_a m_a R_2}{p_a}\right)-\exp\left(-\frac{2 E_\gamma \Gamma_a t_{\rm max}}{m_a}\right)\right]\Theta\left(t_{\rm max}-\frac{m_a^2 r_{\rm prog}}{2E_a E_\gamma v_a}\right).
\end{equation}
Also here we expect that, for the lowest couplings that can be constrained, the exponential can be expanded to give
\begin{equation}
    \left(\frac{d\mathcal{F}_\gamma}{dE_\gamma}\right)_{a\to\gamma\gamma}=\int dE_a \frac{g_{a\gamma}^2 m_a^2 E_\gamma t_{\rm max}}{64\pi^2 r^2  p_a}\frac{dN_a}{dE_a}.
\end{equation}
Hence, the emissivity grows in proportion to $g_{a\gamma}^2 m_a^2 g_{ae}^2$; since for $m_a\gg 2 m_e$ we have $g_{a\gamma}\propto g_{ae}$, we naturally expect the constraints to follow the scaling $g_{ae}\propto m_a^{-1/2}$. At masses $m_a\ll 2 m_e$, the relation between $g_{a\gamma}$ and $g_{ae}$ is different from that used in the main text
\begin{equation}
    g_{a\gamma}=\frac{\alpha g_{ae}}{\pi m_e}\left[1-\frac{4m_e^2}{m_a^2}\mathrm{arcsin}^2\left(\frac{m_a}{2m_e}\right)\right]\sim -\frac{\alpha g_{ae}}{\pi m_e} \frac{m_a^2}{12 m_e^2}.
\end{equation}
Hence, in this case, the emissivity grows in proportion to $g_{ae}^4 m_a^6$, so that the bounds scale as $g_{ae}\propto m_a^{-3/2}$; while we do recover this scaling numerically, we do not show results in this region of parameter space which is, in any event, excluded by cosmology~\cite{Langhoff:2022bij}.

\end{widetext}

\bibliographystyle{bibi}
\bibliography{References.bib}

\end{document}